\documentclass[num-refs]{amse-new}

\newcommand{\be}{\begin{equation}}
\newcommand{\ee}{\end{equation}}
\newcommand{\beaa}{\begin{eqnarray*}}
\newcommand{\eeaa}{\end{eqnarray*}}
\newcommand{\bea}{\begin{eqnarray}}
\newcommand{\eea}{\end{eqnarray}}
\newcommand{\lbl}{\label}
\def\proof {{\noindent\bf Proof.}\quad}
\def\var{\mathrm {var}}

\def\tr{\mathrm {tr}}

\def\w{w}
\def\A{{A}}

\def\z{{z}}

\def\S{{S}}
\def\R{{R}}
\def\I{{I}}

\def\bmu{{\mu}}

\def\X{{X}}

\def\Y{{Y}}

\def\W{{W}}

\def\O{{\Omega}}
\def\D{{D}}

\def\tr{\mathrm {tr}}

\def\bms{{\Sigma}}

\def\cd{\mathop{\rightarrow}\limits^{d}}

\def\bmv{\varepsilon}

\def\boxit#1{\vbox{\hrule\hbox{\vrule\kern6pt\vbox{\kern6pt#1\kern6pt}\kern6pt\vrule}\hrule}}

\def\proof {{\noindent\bf Proof.}\quad}
\def\var{\mathrm {var}}

\def\tr{\mathrm {tr}}

\def\A{{A}}

\def\z{{z}}

\def\S{{S}}
\def\R{{R}}
\def\I{{I}}

\def\bmu{{\mu}}

\def\X{{X}}

\def\Y{{Y}}

\def\W{{W}}

\def\O{{\Omega}}
\def\D{{D}}

\def\tr{\mathrm {tr}}

\def\bms{{\Sigma}}

\def\cd{\mathop{\rightarrow}\limits^{d}}

\def\bmv{\varepsilon}

\def\boxit#1{\vbox{\hrule\hbox{\vrule\kern6pt\vbox{\kern6pt#1\kern6pt}\kern6pt\vrule}\hrule}}

\numberwithin{equation}{section} 

\begin{document}

 \PageNum{1}
 \Volume{202x}{Sep.}{x}{x}
 \OnlineTime{August 15, 202x}
 \DOI{0000000000000000}
 \EditorNote{Received x x, 202x, accepted x x, 202x\\
 The research of Dachuan Chen is supported by the National Natural Science Foundation of China (Grants 12101335 and 12271271), the Natural Science Foundation of Tianjin (Grant 21JCQNJC00020) and the Fundamental Research Funds for the Central Universities, Nankai University (Grants 63211088 and 63221050). Decai Liang was supported by National Natural Science Foundation of China Grant 12101332. Long Feng was partially supported by Shenzhen Wukong Investment Company, the Fundamental Research Funds for the Central Universities under Grant No. ZB22000105, the China National Key R\&D Program (Grant Nos. 2019YFC1908502, 2022YFA1003703, 2022YFA1003802, 2022YFA1003803) and the National Natural Science Foundation of China Grants (Nos. 12271271, 11925106, 12231011,11931001 and 11971247).\\
$^\dagger$ We are both co-corresponding authors and have made equal contributions to this paper.
 }

\abovedisplayskip 6pt plus 2pt minus 2pt \belowdisplayskip 6pt
plus 2pt minus 2pt
\def\vsp{\vspace{1mm}}
\def\th#1{\vspace{1mm}\noindent{\bf #1}\quad}
\def\proof{\vspace{1mm}\noindent{\it Proof}\quad}
\def\no{\nonumber}
\newenvironment{prof}[1][Proof]{\noindent\textit{#1}\quad }
{\hfill $\Box$\vspace{0.7mm}}
\def\q{\quad} \def\qq{\qquad}
\allowdisplaybreaks[4]


\AuthorMark{CHEN, LIANG and FENG}                             

\TitleMark{Asymptotic Independence}  

\title{Asymptotic Independence of the Quadratic form and Maximum of Independent Random Variables with Applications to High-Dimensional Tests}                  

\author{Dachuan \uppercase{Chen}}             
    {Address: School of Statistics and Data Science, KLMDASR, LEBPS, and LPMC, Nankai University, Tianjin, 300071, CHINA\\
    E-mail\,$:$ dchen@nankai.edu.cn}

\author{Decai \uppercase{Liang}$^\dagger$}             
    {Address: School of Statistics and Data Science, KLMDASR, LEBPS, and LPMC, Nankai University, Tianjin, 300071, CHINA\\
    E-mail\,$:$ liangdecai@nankai.edu.cn}

\author{Long FENG$^\dagger$}     
    {Address: School of Statistics and Data Science, KLMDASR, LEBPS, and LPMC, Nankai University, Tianjin, 300071, CHINA\\
    E-mail\,$:$  flnankai@nankai.edu.cn}

\maketitle%

\Abstract{This paper establishes the asymptotic independence between the quadratic form and maximum of a sequence of independent random variables. Based on this theoretical result, we find the asymptotic joint distribution for the quadratic form and maximum, which can be applied into the high-dimensional testing problems. By combining the sum-type test and the max-type test, we propose the Fisher's combination tests for the one-sample mean test and two-sample mean test. Under this novel general framework, several strong assumptions in existing literature have been relaxed. Monte Carlo simulation has been done which shows that our proposed tests are strongly robust to both sparse and dense data.}      


\MRSubClass{62F03, 62H15}      


\section{Introduction}

Independence is a very important property in statistical inference. In this paper, we develop the asymptotic independence between the quadratic form $\z^\top\A\z$ and the maximum $\max_{1\le i\le p} |z_i|$ of a sequence of independent sub-Gaussian random variables $\z=(z_1,\cdots,z_p)^{\top}$, where $\A$ is a symmetric matrix. The benefits of this theoretical result will be reflected in the application of high-dimensional tests, including the one-sample mean test and two-sample mean test.

Yet little research has been done on the asymptotic joint distribution between the quadratic form and the maximum of a sequence random variables. This is the first paper on this topic. In contrast, the majority of the existing literature are focusing on the development of the asymptotic independence between the sum $\sum_{i=1}^{p} X_i$ and the maximum $\max_{1 \le i \le p} X_i$ of a sequence of random variables $\{X_i\}_{i=1}^{p}$. Here we provide a brief review for the related literature. \cite{chow1978sum} derives the asymptotic independence between the sum and maximum by assuming $\{X_i\}_{i=1}^{p}$ to be independent and identically distributed (i.i.d., hereafter). There are two main streams of statistical work which have relaxed the i.i.d. assumption. On one hand, \cite{anderson1991joint,anderson1993limiting,anderson1995sums} and \cite{hsing1995note} extends the theoretical results by assuming the random variables $\{X_i\}_{i=1}^{p}$ satisfying the strong mixing condition. On the other hand, \cite{ho1996asymptotic}, \cite{ho1999asymptotic}, \cite{mccormick2000asymptotic} and \cite{peng2003joint} establishes the asymptotic independence between the sum and the maximum based on the assumption that $\{X_i\}_{i=1}^{p}$ is a stationary Gaussian sequence. \cite{feng2022test} further investigates the asymptotic independence between the sum and the maximum of the squares of the dependent random variables without imposing the stationary assumption.

In this paper, the asymptotic joint distribution of the quadratic form $\z^\top\A\z$ and the maximum $\max_{1\le i\le p} |z_i|$ of independent sub-Gaussian random variables has been derived, where the random variables $\{z_i\}_{i=1}^p$ are assumed to have mean zero and variance one. The asymptotic normality of the standardized quadratic form $\frac{\z^\top\A\z-\tr(\A)}{\sigma_A}$ is guaranteed by the assumption that $\A$ is symmetric and $\sup _{i} \sum_{j=1}^{p}\left|a_{i j}\right|<K$. The asymptotic distribution of the maximum $\max_{1\le i\le p} |z_i|$ is derived based on the assumption that $pP(|z_i|>l_p(y))\to h(y)$ for all $i$ where $l_p(y)\to \infty$ and $h(y)$ is bounded. The asymptotic independence between the standardized quadratic form and the maximum is mainly based on the assumption that the smallest eigenvalue of $\A$ is bounded away from zero and the largest eigenvalue of $\A$ is bounded.

Our theoretical result on asymptotic independence is novel and different from the existing ones. As a consequence, this general framework provides a new way for us to look at the theoretical foundation of the high-dimensional testing problems, including the one-sample mean test and the two-sample mean test.

The research of high-dimensional hypothesis tests has been evolved rapidly in the last two decades, which has been widely applied to a range of areas, including genomics, neuroscience, finance, economics and so on. In general, high-dimensionality means that the data dimension can be larger than the sample size, or the data dimension can also grow to infinity in asymptotics. Under this dimension setting, the classical statistical testing theories no longer applicable. For example, the traditional Hotelling's $T^2$ test cannot work when the data dimension exceeds the sample size because of the singularity of sample covariance matrix. Consequently, the high dimensional testing problems are grounded on a new theoretical foundation comparing with that of the classical ones. By replacing the sample covariance matrix in Hotelling's $T^2$ test with the identity matrix or the diagonal matrix of the sample covariance matrix, several sum-type tests has been proposed for the high-dimensional mean test problem, see e.g., \cite{bai1996effect}, \cite{srivastava2008test}, \cite{srivastava2009test}, \cite{chen2010two} and \cite{srivastava2013two}. However, due to the low performance of the sum-type test under the sparse alternative, where there are a few nonzero elements in the mean vector or the mean difference, many efforts have been made to improve the sum-type tests, see, for example, \cite{zhong2013tests}, \cite{fan2015power} and \cite{chen2019two}. Different from the sum-type tests, the other solution for sparse alternative is the max-type tests proposed by \cite{tony2014two}, which is particularly powerful for the sparse data but cannot work well with the dense data.

In real world, it is usually difficult to identify whether the data is sparse or not. Therefore, it becomes necessary to develop a test which can work well for both sparse and dense alternatives. The power enhancement test proposed by \cite{fan2015power} is one candidate for this purpose which adding a screening statistic to the sum-type test statistic. The other candidate is the adaptive test proposed by \cite{xu2016adaptive} which studied the asymptotic independence between the max-type test and the sum-of-powers tests and then combined them together based on their p-values. \cite{he2021asymptotically} is another candidate for the same purpose which combined the max-type test with a set of finite-order U-Statistics based on their asymptotic independence. Also note that for testing the high-dimensional covariance matrices, \cite{yu2020fisher, yu2022power} studied the Fisher's combination test for asymptotic independent statistics. In this paper, based on the novel theoretical result on the asymptotic independence between the quadratic form and maximum of independence random variables, the aforementioned problem has been solved, and at the same time, several strong assumptions made by \cite{xu2016adaptive} and \cite{he2021asymptotically} have been relaxed. We propose two Fisher's combination tests by combining the max-type tests and sum-type tests for the one-sample mean test and two-sample mean test. The simulation results show that our proposed tests are strongly robust to both sparse data and dense data.

The main contributions of this paper are listed as follows:
\begin{enumerate}
\item We show the asymptotic independence between the quadratic form and the maximum of independent sub-Gaussian random variables, which is novel in existing literature;
\item Based on the above theoretical results, we have proposed the Fisher's combination test by combining the sum-type test and the max-type test for two types of high-dimensional testing problems: one-sample mean test and two-sample mean test. Simulation results show that our proposed tests are robust to both sparse and dense data;
\item  The development of these two applications reflects the theoretical benefits of our general framework in proving the asymptotic independence between the max-type tests and sum-type tests: (1) the strong assumptions on the population covariance structure (i.e., the $\alpha$-mixing condition or the diagonal assumption) in existing literature have been relaxed; (2) besides the sub-Gaussian-type tails, our theoretical development also allows the polynomial-type tail for the sample distribution;
\item By switching the alternative hypothesis to the special local alternative, for example, sparse transformed mean in one-sample mean test or sparse transformed mean difference in two-sample mean test, we could obtain the asymptotic independence between the max-type test and the sum-type test under the alternative hypothesis. As a consequence, the expression of the power of our proposed test has been derived, which is the first result on this topic.
\end{enumerate}

The organization of this paper is as follows. In Section 2, we provide the basic definition about the distribution of the random variables and state the theoretical result about the asymptotic independence between the quadratic form and maximum of the independent random variables. In Section 3, we apply the theoretical result into two types of tests of high dimensional data. In Section 4, the proposed tests are compared with some existing ones via Monte Carlo simulation. The mathematical proofs of our theoretical results are collected in the online supplementary material.

\section{Asymptotic Independence of the Quadratic form and Maximum of Independent Random Variables}

In this section, we provide the central theoretical results: the asymptotic independence between the quadratic form and maximum of independent sub-Gaussian random variables. The statement will start with the definition of the sub-Gaussian random variable.

\begin{definition}
A random variable $X$ with mean $\mu=\mathbb{E}[X]$ is $\sigma^2$-sub-Gaussian if there is a positive number $\sigma$ such that
\begin{align} \label{sb}
\mathbb{E}\bigg[e^{\lambda(X-\mu)}\bigg] \leq e^{\sigma^{2} \lambda^{2} / 2} \quad \text { for all } \lambda \in \mathbb{R}
\end{align}
\end{definition}
The constant $\sigma$ is referred to as the sub-Gaussian parameter; for instance, we say that $X$ is $\sigma^2$-sub-Gaussian when the condition (\ref{sb}) holds. Naturally, any Gaussian variable with variance $\sigma^{2}$ is sub-Gaussian with parameter $\sigma$.

\begin{theorem}\label{ind}
Assume $z_1,\cdots,z_p$ are independent $\sigma^2$-sub-Gaussian random variables with $E(z_i)=0$ and $\var(z_i)=1$. Suppose the following assumptions hold: (i) There exist two parameters $l_p(y)$ and $h(y)$ satisfies $pP(|z_i|>l_p(y))\to h(y)$ for all $i$ where $l_p(y)\to \infty$ and $h(y)$ is bounded; (ii) $\A$ is symmetric and $\sup _{i} \sum_{j=1}^{p}\left|a_{i j}\right|<K$; (iii) There exist a constant $c>0$ satisfies $c^{-1}<\lambda_{\min}(\A)\le \lambda_{\max}(\A)<c$. Then, we have
\begin{align}\label{th11}
P\bigg(\frac{\z^\top\A\z-\tr(\A)}{\sigma_A}\le x,\max_{1\le i\le p} |z_i| \le l_p(y)\bigg)\to \Phi(x)F_h(y)
\end{align}
where $\sigma_A^2={2\tr(\A^2)+\sum_{i=1}^p a_{ii}^2(E(z_i^4)-3)}$ and $F_h(y)=e^{-h(y)}$.
\end{theorem}

Theorem \ref{ind} showed the asymptotic joint distribution of the standardized quadratic form and the maximum of independent sub-Gaussian random variables. Specifically, the assumption (i) is used to derive the asymptotic distribution of the maximum; the assumption (ii) guarantees the asymptotic normality of the standardized quadratic form; and finally the asymptotic independence between the standardized quadratic form and the maximum is mainly based on assumption (iii).

\begin{remark}
It is worth to mention that if all $|z_i|$'s in Theorem \ref{ind} and in its proof are replaced by $z_i$, the updated theorem and proof still hold. That is, we could show the asymptotic independence between the quadratic form $\frac{\z^\top\A\z-\tr(\A)}{\sigma_A}$ and the maximum $\max_{1\le i\le p} z_i$ based on the similar framework as that of Theorem \ref{ind}.
\end{remark}

\begin{remark}
Note that the existing literature, e.g., \cite{xu2016adaptive}, \cite{yu2020fisher} and \cite{he2021asymptotically}, are all focusing on the derivation of the asymptotic independence between $\max_{1\le i\le p} z_i^2$ and $\sum_{i=1}^{p} z_i^2$. In contrast, this paper developed the asymptotic independence between $\max_{1\le i\le p} z_i$  and $\z^\top\A\z$.
\end{remark}

\section{High Dimensional tests}

In this section, we apply the theoretical results obtained in Section 2 into two types of high-dimensional hypothesis tests: one-sample mean test and two-sample mean test.

\subsection{One-sample problem}
Let $ {X}_{1}, \cdots,  {X}_{n}$ with $ {X}_{i}=\bigg(X_{i 1}, \cdots, X_{i p}\bigg)^{\top}$ for each $i \in\{1, \cdots, n\}$ be a sequence of $p$ dimensional independent and identically distributed (iid) observations from a multivariate distribution with mean vector $\bmu$ and covariance matrix $ \bms$. Our interest is in testing the one sample mean hypotheses
\begin{align}\label{hy1}
H_{0}:  {\mu}=\mathbf{0} \text { versus } H_{1}:  {\mu} \neq \mathbf{0}
\end{align}

The hypothesis test on population mean is a classic and important topic in multivariate statistics, which has been developed in a large statistical literature, see, e.g., \cite{anderson1962introduction}, \cite{eaton1983multivariate} and \cite{muirhead2009aspects} for classical theory. The most famous methodology is the Hotelling's $T^2$ test, see \cite{hotelling1931}. In asymptotics, these classical theories assume the dimension $p$ to be fixed as the sample size $n$ goes to infinity. However, when the dimension $p$ is larger than the sample size $n$, these earlier methods cannot really work. For example, the Hotelling's $T^2$ test requires the inverse of the sample covariance matrix. Actually, the sample covariance matrix is non-invertible when $p>n$. To overcome this difficulty, many high-dimensional mean tests have been proposed by allowing $p \ge n$ and letting both $n$ and $p$ go to infinity in asymptotics.

In this subsection, we focus on the hypothesis testing problem in (\ref{hy1}) under the setting of $p \ge n$. \cite{srivastava2008test} developed the sum-type test statistics for the multivariate normal observations while \cite{srivastava2009test} showed that this sum-type test statistics can also work for the non-normal observations under some certain moment assumptions. In general, as proposed in \cite{srivastava2009test}, the sum-type test statistics can be expressed as follows:
\begin{align}\label{ts1}
T_{SR}=\frac{n \bar{\X}^\top \D_{s}^{-1}\bar{\X}-(n-1) p /(n-3)} {\bigg[2 \operatorname{tr} \hat\R^{2}-p^{2} / (n-1)\bigg]^{\frac{1}{2}}\bigg[1+\bigg(\operatorname{tr}\hat \R^{2}\bigg) / p^{\frac{3}{2}}\bigg]^{\frac{1}{2}}}
\end{align}
where
$$
\begin{aligned}
&\hat\R=\D_{s}^{-\frac{1}{2}} \S \D_{s}^{-\frac{1}{2}},\D_{s}=\operatorname{diag}\bigg(s_{11}, \ldots, s_{p p}\bigg), \\
&\bar{\X}=\frac{1}{n}\sum_{i=1}^n \X_i, \quad \S=\bigg(s_{i j}\bigg)_{1\le i,j\le p}=\frac{1}{n-1}\sum_{i=1}^n(\X_i-\bar{\X})(\X_i-\bar{\X})^\top.
\end{aligned}
$$
As shown in \cite{srivastava2009test}, $T_{SR}$ can be rewritten as follows:
\begin{align*}
T_{SR}=\frac{\w\bms^{1/2}\D^{-1}\bms^{1/2}\w-(n-1) p /(n-3)}{\sqrt{2\tr(\R^2)}}+o_p(1)
\end{align*}
where $\w=(w_1,\cdots,w_p)^\top, w_i=n^{-1/2}\sum_{k=1}^n\varepsilon_{ki}$ under the following condition (C1). So we observe that $T_{SR}$ asymptotically has a quadratic form of a sequence of independent random variables.

As pointed out by \cite{tony2014two}, the aforementioned sum-type test cannot work well when the mean vector is sparse, for example, there are a few nonzero elements in the mean vector. To improve the sum-type test under the sparse alternative, \cite{zhong2013tests} established a new test by thresholding two sum-type tests based on the sample means and then maximizing over a range of thresholding levels. The other effort is made by \cite{fan2015power}, which proposed a linear combination between the standard Wald statistic and a power enhancement component as the test statistic, where the power enhancement component equals zero with probability converging to one under null and diverges in probability under some specific regions of alternatives. In contrast, the max-type test statistics proposed by \cite{tony2014two} might be most powerful under the sparse alternative. For convenience, we write the expression of the max-type test statistic for the one-sample mean test as follows.

For a given invertible $p \times p$ matrix $ {A}$, the null hypothesis $H_{0}:  \bmu=0$ is equivalent to $H_{0}:  {A}  \bmu=0 . \operatorname{Set}   \delta^{ {A}}=\bigg(\delta_{1}^{ {A}}, \ldots, \delta_{p}^{ {A}}\bigg)^{\prime}:= {A}\bar{\X}$ . Denote the sample covariance matrix of $ {A X}$ by $ {B}=\bigg(b_{i, j}\bigg)$ and define the test statistic
\begin{align} \label{DEF_max}
M_{ {A}}=(n-1) \max _{1 \leqslant i \leqslant p} \frac{\bigg(\delta_{i}^{ {A}}\bigg)^{2}}{b_{i, i}}.
\end{align}
Similar to \cite{tony2014two}, we proposed the following max-type test statistic $M_{\hat{\O}^{1/2}}$ by choosing $\A$ as $\hat{\O}^{1/2}$. Here $\hat{\O}$ is a good estimator of the precision matrix $\O= \bms^{-1}$. Under the following condition (C2), we can rewrite $M_{\hat{\O}^{1/2}}$ as
\begin{align*}
M_{\hat{\O}^{1/2}}=\max_{1\le i\le p} w_i^2+o_p(1)
\end{align*}
which is a maximum of a sequence of independent random variables.

In what follows, we state the required assumptions for the development of the asymptotic distribution of the max-type test statistic $M_{\hat{\O}^{1/2}}$.

\begin{itemize}

\item[(C1)] $\X_i=  \mu+ \bms^{1/2}  \varepsilon_i$ where $  \varepsilon_i=(\varepsilon_{i1},\cdots,\varepsilon_{ip})^\top$ and $\varepsilon_{ij}$ are independently distributed with $E(\varepsilon_{ij})=0, \var(\varepsilon_{ij})=1$ and $E(\varepsilon_{ij}^4)<c$ for some positive constant $c$.
\item[(C2)]  $C_{0}^{-1} \leqslant \lambda_{\min }( {\Sigma}) \leqslant \lambda_{\max }( {\Sigma}) \leqslant C_{0}$ for some constant $C_{0}>0 .$
\item[(C3)] We assume that the estimator $\hat{\O}=(\hat{\omega}_{ij})$ has at least a logarithmic rate of convergence
\begin{align*}
\|\hat{\Omega}-\Omega\|_{L_{1}}=o_{{p}}\bigg\{\frac{1}{\log (p)}\bigg\},\max _{1 \leqslant i \leqslant p}\left|\hat{\omega}_{i, i}-\omega_{i, i}\right|=o_{p}\bigg\{\frac{1}{\log (p)}\bigg\}
\end{align*}
\item[(C4)] Suppose the following condition (i) or (ii) hold: (i)(sub-Gaussian-type tails) There exist some constant $\eta>0, K>0$ such that $E(\exp(\eta \varepsilon_{ij}^2))\le K$. And $\log p=o(n^{1/4})$; (ii) (polynomial-type tails) Suppose that for some constants $\gamma_0, c_1>0, p\le c_1 n^{\gamma_0}$ and for some positive constant $\epsilon, K$, $E(|\varepsilon_{ij}|^{2\gamma_0+2+\epsilon})\le K$.
\end{itemize}

The limiting null distribution of the max-type test statistic $M_{{\hat\O}^{1/2}}$ is given in following theorem.

\begin{theorem}\label{th2}

Suppose that conditions (C1)-(C4) hold. For any $x \in \mathbb{R}$,
$$
P_{H_{0}}\bigg[M_{{\hat\O}^{1/2}}-2 \log (p)+\log \{\log (p)\} \leqslant x\bigg] \rightarrow \exp \bigg\{-\frac{1}{\sqrt{\pi}} \exp \bigg(-\frac{x}{2}\bigg)\bigg\}, \quad \text { as } p \rightarrow \infty
$$

\end{theorem}

Given the above description, we know that the sum-type test can only work well under the dense alternative, while the max-type test can only work well under the sparse alternative. In order to achieve good performance under both sparse and dense alternatives, it is a good choice to propose a combination of the sum-type test and the max-type test based on their asymptotic independence. The existing literature on this combination approach includes \cite{xu2016adaptive} and \cite{he2021asymptotically}. \cite{xu2016adaptive} showed the asymptotic independence between the max-type test and a family of the sum-of-powers tests and constructed the combined test by picking up the minimum of the p-values of these tests in order to reach the maximum power. \cite{he2021asymptotically} developed the asymptotic independence between the max-type test statistics and the finite-order U-statistics and combined these tests based on the minimum p-values or the Fisher's method, see e.g., \cite{littell1971asymptotic}.

Compared with these existing methodology, our theoretical contribution is that several strong assumptions in existing literature have been relaxed. First, both \cite{xu2016adaptive} and \cite{he2021asymptotically} only assume the sub-Gaussian-type tail for the sample distribution. In contrast, our theoretical framework also allows for the polynomial-type tail, e.g., see condition (C4). Second, we impose weaker assumptions for the covariance structure than those of \cite{xu2016adaptive} and \cite{he2021asymptotically}. The detailed assumption and discussion are provided as follows.

In the following condition, we state the additional assumption about the covariance structure, which is required to show the asymptotic independence between the sum-type test and max-type test.

\begin{itemize}
\item[(C5)] $\sup_{i}\sum_{j=1}^p |a_{ij}|<K$ where $\A =  \bms^{1/2} \D^{-1}  \bms^{1/2} =(a_{ij})_{1\le i,j\le p}$ and $\D$ is the diagonal matrix of $ \bms$.
\end{itemize}

In existing literatures related to the asymptotic independence between max-type tests and sum-type tests, the assumptions about covariance structure are relatively strong. For example, the Theorem 1 of \cite{xu2016adaptive} is based on the assumption of $\alpha$-mixing condition. Moreover, the Theorem 2.3 of \cite{he2021asymptotically} assumed the covariance structure to be diagonal. In contrast, our assumption about the covariance structure (as stated in condition (C5)) is more general. According to condition (C2), the eigenvalues of $\A$ are also bounded. So the conditions (ii) and (iii) in Theorem \ref{ind} hold. How to relax the bounded eigenvalues assumption of $A$ deserves some further studies.

Up to now, based on the result in Theorem \ref{ind}, we could show the asymptotic independence between $T_{SR}$ and $M_{{\hat\O}^{1/2}}$ under the null hypothesis.

\begin{theorem}\label{th3}
Suppose that conditions (C1)-(C5) hold. For any $x,y \in \mathbb{R}$,
\begin{align}
P_{H_{0}}\bigg[T_{SR}\le x,M_{{\hat\O}^{1/2}}-2 \log (p)+\log \{\log (p)\} \leqslant y\bigg] \rightarrow \Phi(x)F(y)
\end{align}
where $F(y)=\exp(-\frac{1}{\sqrt{\pi}}e^{-y/2})$.
\end{theorem}
To combine the proposed max-type and sum-type tests, we propose the
Fisher's combination test,
based on the asymptotic independence between $T_{SR}$
and $M_{{\hat\O}^{1/2}}$.
Specifically, let $$p_{\operatorname{MAX}}^{(1)}\doteq
1-F\bigg\{M_{{\hat\O}^{1/2}}-2 \log (p)+\log \{\log (p)\}\bigg\}\operatorname{  and }p_{\operatorname{SUM}}^{(1)}\doteq
1-\Phi\bigg(T_{SR}\bigg)$$
denote the $p$-values with respect to the test statistics
$T_{\operatorname{MAX}}^{(1)}=M_{{\hat\O}^{1/2}}$ and $T_{\operatorname{SUM}}^{(1)}=T_{SR}$
respectively. Based on $p_{\operatorname{MAX}}^{(1)}$ and
$p_{\operatorname{SUM}}^{(1)}$, the proposed Fisher's combination test
rejects $H_0$ at the significance level $\alpha$, if
\begin{align}\label{DEF_T_FC}
T_{\textrm{FC}}^{(1)}\doteq -2 \log p_{\operatorname{MAX}}^{(1)}-2\log p_{\operatorname{SUM}}^{(1)}
\end{align}
is larger than $c_{\alpha}$, i.e. the $1-\alpha$ quantile of the
chi-squared distribution with 4 degrees of freedom.

Based on Theorem \ref{th3}, we immediately have the following result
for $T_{\textrm{FC}}^{(1)}$.

\begin{corollary}
Assume the same conditions as in Theorem \ref{th3}, then we have
$T_{\textrm{FC}}^{(1)}\cd \chi_4^2$ as $n,p\to \infty$.
\end{corollary}

We consider the following alternative hypothesis:
\begin{align}\label{h12}
H_1: \tilde\mu_i\not=0, i\in \mathcal{M}, ~~ |\mathcal{M}|=m, ~~m=o(p^{1/2}), ~~   \mu=\frac{  \delta}{(np)^{1/2}}
\end{align}
where $\tilde{  \mu}=\O^{1/2}  \mu$ and $  \delta$ is a vector of constants and $  \delta^\top \D^{-1}  \delta\le Cp$ for some constant $C$.

It is easy to see that the local alternative $H_1$ in (\ref{h12}) is a special case of the original alternative hypothesis $H_1:  {\mu} \neq  {0}$. Under this special local alternative, we could further obtain the asymptotic independence between the sum-type test and max-type test as follows.

\begin{theorem}\label{th4}
Suppose that conditions (C1)-(C5) hold. Under the special local alternative $H_{1}$ stated in (\ref{h12}), for any $x,y \in \mathbb{R}$,
\begin{align*}
&P\bigg[T_{SR}\le x,M_{{\hat\O}^{1/2}}-2 \log (p)+\log \{\log (p)\} \le y\bigg] \\
&\rightarrow P\bigg[T_{SR}\le x\bigg] P\bigg[M_{{\hat\O}^{1/2}}-2 \log (p)+\log \{\log (p)\} \le y\bigg].
\end{align*}
\end{theorem}

Based on Theorem \ref{th4}, we can analysis the power function of our proposed Fisher Combination test under the above special alternative hypothesis (\ref{h12}).
Define a minimal p-values test $T_{\min}^{(1)}=\min(p_{\operatorname{SUM}}^{(1)},p_{\operatorname{MAX}}^{(1)})$. According to Theorem \ref{th3}, we reject the null hypothesis if $p_{\operatorname{SUM}}^{(1)}\le 1-\sqrt{1-\gamma}\approx \gamma/2$ or $p_{\operatorname{MAX}}^{(1)}\le 1-\sqrt{1-\gamma}\approx \gamma/2$.
According to the results in \cite{littell1971asymptotic,littell1973asymptotic}, we have that the power of Fisher combination test is asymptotically optimal in terms of Bahadur relative efficiency. In simulations, we found that the power of $\beta_{T_{FC}^{(1)}}$ is larger than the power of the minimal p-values test $\beta_{T_{\min}^{(1)}}$ in most cases. We also have
\begin{align*}
\beta_{T_{\min}^{(1)}}\ge\beta_{T_{\operatorname{SUM}}^{(1)}, \gamma/2}+\beta_{T_{\operatorname{MAX}}^{(1)},\gamma/2}-\beta_{T_{\operatorname{SUM}}^{(1)}, \gamma/2}\beta_{T_{\operatorname{MAX}}^{(1)},\gamma/2}
\end{align*}
where the last inequality is based on the inclusion-exclusion principle and the result of Theorem \ref{th4}, and $\beta_{T_{\operatorname{SUM}}^{(1)}, \gamma}$ is the power function of the sum-type test $T_{\operatorname{SUM}}^{(1)}$ at significant value $\gamma$. So does $\beta_{T_{\operatorname{MAX}}^{(1)},\gamma}$.

\subsection{Two-sample problem}

Assume that
$\{\X_{i1},\cdots,\X_{in_i}\}$ for $i=1,2$ are two independent
random samples with the sizes $n_1$ and $n_2$, from $p$-variate
distributions $F({\bf x}- \bmu_1)$ and $G({\bf x}- \bmu_2)$ located at
$p$-variate centers $ \bmu_1$ and $ \bmu_2$. Let $n=n_1+n_2$. We
wish to test
\begin{align}\label{ht}
H_0: \bmu_1= \bmu_2\ \ \mbox{versus}\ \  H_1: \bmu_1\neq \bmu_2,
\end{align}
when their common covariances $ \bms$ is unknown.

The famous Hotelling's $T^2$ test statistic for the two-sample problem also requires the dimension $p$ to be fixed in asymptotics. Because it relies on the inverse of the pooled sample covariance matrix, Hotelling's $T^2$ test cannot work when $p \ge n$. To deal with the problem of high-dimensional setting, i.e., when $p \ge n$, several sum-type tests have been proposed. \cite{bai1996effect} proposes a test by replacing the pooled sample covariance matrix in Hotelling's $T^2$ test statistic with the identity matrix. \cite{srivastava2008test} developed a test for two set of multivariate normal observations which sharing the same population covariance matrix. Under some mild assumptions on the covariance structure, \cite{chen2010two} derived a test statistic and relaxed the assumption $p/n \to c \in (0, \infty)$ in \cite{bai1996effect} by allowing the arbitrarily large $p$ which can be independent with the sample size $n$. For the case of unequal covariance matrix, \cite{srivastava2013two} proposed the following sum-type test statistics:
\begin{align} \label{DEF_T_SKK}
T_{SKK}=\frac{\bigg(\bar{\X}_{1}-\bar{\X}_{2}\bigg)^{\top} \hat{\D}^{-1}\bigg(\bar{\X}_{1}-\bar{\X}_{2}\bigg)-p}{\sqrt{p \hat{\sigma}_{SKK}^2c_{p, n}}}
\end{align}
where
\begin{align*}
\hat{\sigma}_{SKK}^2=&\frac{2\tr(\hat{\R}^2)}{p}-\frac{2}{p(n_1-1)n_1^2}\bigg(\tr(\hat{\D}^{-1}\S_1)\bigg)^2-\frac{2}{p(n_2-1)n_2^2}\bigg(\tr(\hat{\D}^{-1}\S_2)\bigg)^2,\\
c_{p, n}=&1+\frac{\operatorname{tr} \hat{\R}^{2}}{p^{3 / 2}}, \bar{\X}_{i}=\frac{1}{n_i}\sum_{l=1}^{n_i} \X_{il} \mbox{ for } i=1,2
\end{align*}
where $\S_1$ and $\S_2$ are the sample covariance matrices of $\{\X_{1l}\}_{l=1}^{n_1}$ and $\{\X_{2l}\}_{l=1}^{n_2}$, respectively. $\hat\D=\frac{1}{n_1}\hat{\D}_1+\frac{1}{n_2}\hat{\D}_2$ where $\hat{\D}_i, i=1,2$ is the diagonal matrix of $\S_i$. And $\hat{\R}=\hat{\D}^{-1/2}\bigg(\frac{1}{n_1}\hat{\S}_1+\frac{1}{n_2}\hat{\S}_2\bigg)\hat{\D}^{-1/2}$. Under condition (C1$'$), we can also rewrite $T_{SKK}$ as a quadratic form of a sequence of independent random variables, i.e.
\begin{align*}
T_{SKK}=\frac{1}{\sqrt{2\tr(\R^2)}}\bigg(  u^\top \A  u-p\bigg)+o_p(1)
\end{align*}
where $\A$ is defined in condition (C5) and $  u=\sqrt{\frac{n_1n_2}{n_1+n_2}}\bigg(\frac{1}{n_1}\sum_{l=1}^{n_1}  \varepsilon_{1l} - \frac{1}{n_2}\sum_{l=1}^{n_2}  \varepsilon_{2l}\bigg)$.

Similar to the discussion in the one-sample problem, the aforementioned sum-type tests cannot work well under the sparse alternatives. Here, sparse alternative means that the difference of the population means is sparse. To overcome this difficulty, \cite{chen2019two} proposed a test based on the thresholding technique and data transformation, which can be regard as the extension of the method in \cite{zhong2013tests}. For the sparse alternative, \cite{tony2014two} proposed the following max-type test statistics:
\begin{align}\label{max2}
W_{\hat{\O}^{1/2}}=\frac{n_1n_2}{n_1+n_2}\max_{1\le i\le p}\bar{W}_i^2
\end{align}
where $\bar{\W}=(\bar{W}_1,\cdots,\bar{W}_p)\doteq \hat{\O}^{1/2}(\bar{\X}_1-\bar{\X}_2)$. Similarly, we can also rewrite $W_{\hat{\O}^{1/2}}$ as a maximum of a sequence of independent random variables, i.e.
\begin{align*}
W_{\hat{\O}^{1/2}}=\max_{1\le i\le p} u_i^2+o_p(1)
\end{align*}
where $u=(u_1,\cdots,u_p)^\top$.

To facilitate the description of theoretical results in two-sample mean test, we switch the condition (C1) in previous section to a new one as follows.

\begin{itemize}

\item[(C1$'$)] For $k=1,2$, $\X_{ki}=  \mu_k+ \bms^{1/2}  \varepsilon_{ki}$ where $  \varepsilon_{ki}=(\varepsilon_{ki1},\cdots,\varepsilon_{kip})$ and $\varepsilon_{kij}$ are independently distributed with $E(\varepsilon_{kij})=0, \var(\varepsilon_{kij})=1$ and $E(\varepsilon_{kij}^4)<c$ for some positive constant $c$.
\end{itemize}

The limiting null distribution of the max-type test statistic $W_{{\hat\O}^{1/2}}$ is as follows.

\begin{theorem}\label{th5}

Suppose that conditions (C1$'$), (C2)-(C4) hold. For any $x \in \mathbb{R}$,
$$
P_{H_{0}}\bigg[W_{{\hat\O}^{1/2}}-2 \log (p)+\log \{\log (p)\} \leqslant x\bigg] \rightarrow \exp \bigg\{-\frac{1}{\sqrt{\pi}} \exp \bigg(-\frac{x}{2}\bigg)\bigg\}, \quad \text { as } p \rightarrow \infty
$$
\end{theorem}

Because it is difficult to tell whether the data is sparse or not in the real world, we need to develop a test which is robust to both sparse and dense alternatives at the same time. The main idea of the solution is to combine the sum-type test and max-type test based on the asymptotic independence between them, which is similar to those of \cite{xu2016adaptive} and \cite{he2021asymptotically}.

To achieve the asymptotic independence between sum-type test and max-type test, we now apply the theoretical results stated in Section 2 into the two-sample mean test as follows.

\begin{theorem}\label{th6}
Suppose that conditions (C1$'$), (C2)-(C5) hold. For any $x,y \in \mathbb{R}$,
\begin{align}
P_{H_{0}}\bigg[T_{SKK}\le x,W_{{\hat\O}^{1/2}}-2 \log (p)+\log \{\log (p)\} \leqslant y\bigg] \rightarrow \Phi(x)F(y)
\end{align}
\end{theorem}

It is clear that in the two-sample problem, the result of asymptotic independence shares the same theoretical benefits as that of the one-sample problem. For example, compared with \cite{xu2016adaptive} and \cite{he2021asymptotically}, our result relies on the weaker assumption on the covariance structure and allows for more possibilities for the assumptions about the sample distributions.

Based on the asymptotic independence between $T_{SKK}$
and $W_{{\hat\O}^{1/2}}$, we propose the Fisher's combination test which utilizing
the max-type and sum-type tests:
\begin{align}\label{DEF_T_FC_2}
T_{\textrm{FC}}^{(2)}\doteq -2 \log p_{\operatorname{MAX}}^{(2)}-2\log p_{\operatorname{SUM}}^{(2)}
\end{align}
where
$$p_{\operatorname{MAX}}^{(2)}\doteq
1-F\bigg\{W_{{\hat\O}^{1/2}}-2 \log (p)+\log \{\log (p)\}\bigg\} \operatorname{ and }p_{\operatorname{SUM}}^{(2)}\doteq
1-\Phi\bigg(T_{SKK}\bigg)$$
denote the $p$-values with respect to the test statistics
$T_{\operatorname{MAX}}^{(2)}=W_{{\hat\O}^{1/2}}$ and $T_{\operatorname{SUM}}^{(2)}=T_{SKK}$
respectively.

Based on Theorem \ref{th6}, we immediately have the following result
for $T_{\textrm{FC}}^{(2)}$.

\begin{corollary}
Assume the same conditions as in Theorem \ref{th6}, then we have
$T_{\textrm{FC}}^{(2)}\cd \chi_4^2$ as $n,p\to \infty$.
\end{corollary}

We consider the following alternative hypothesis:
\begin{align}\label{h22}
H_1: \tilde\mu_i\not=0, i\in \mathcal{M}, ~~ |\mathcal{M}|=m, ~~m=o(p^{1/2}), ~~   \mu_1- \bmu_2=\frac{  \delta}{(np)^{1/2}}
\end{align}
where $\tilde{  \mu}=\O^{1/2}(  \mu_1- \bmu_2)$ and $  \delta$ is a vector of constants and $  \delta^\top \D^{-1}  \delta\le Cp$ for some constant $C$.

As the special case of the original alternative hypothesis $H_1: \bmu_1\neq \bmu_2$, the special local alternative $H_1$ in (\ref{h22}) enables us to obtain the asymptotic independence between the sum-type test $T_{SKK}$ and the max-type test $W_{{\hat\O}^{1/2}}$ under the alternative hypothesis. The main result is stated in the following theorem.

\begin{theorem}\label{th7}
Suppose that conditions (C1$'$), (C2)-(C5) hold. Under the special local alternative $H_{1}$ stated in (\ref{h22}), for any $x,y \in \mathbb{R}$,
\begin{align*}
&P\bigg[T_{SKK}\le x,W_{{\hat\O}^{1/2}}-2 \log (p)+\log \{\log (p)\} \le y\bigg] \\
&\rightarrow P\bigg[T_{SKK}\le x\bigg] P\bigg[W_{{\hat\O}^{1/2}}-2 \log (p)+\log \{\log (p)\} \le y\bigg].
\end{align*}
\end{theorem}

Based on Theorem \ref{th7}, we can analysis the power function of $T_{FC}^{(2)}$ for the two-sample problem.
Similar to the one-sample problem, based on Theorems \ref{th6} and \ref{th7} and the results in \cite{littell1971asymptotic,littell1973asymptotic}, and  by defining the minimal p-values test as $T_{\min}^{(2)}=\min(p_{\operatorname{SUM}}^{(2)},p_{\operatorname{MAX}}^{(2)})$, we have the following relationship among the powers of different tests: $\beta_{T_{FC}^{(2)}}$ is slightly larger than $\beta_{T_{\min}^{(2)}}$ in most cases in our simulation studies.  And
\begin{align*}
\beta_{T_{\min}^{(2)}}\ge\beta_{T_{\operatorname{SUM}}^{(2)}, \gamma/2}+\beta_{T_{\operatorname{MAX}}^{(2)},\gamma/2}-\beta_{T_{\operatorname{SUM}}^{(2)}, \gamma/2}\beta_{T_{\operatorname{MAX}}^{(2)},\gamma/2}
\end{align*}
where $\beta_{T_{\operatorname{SUM}}^{(2)}, \gamma}$ is the power function of the sum-type test $T_{\operatorname{SUM}}^{(2)}$ at significant value $\gamma$. So does $\beta_{T_{\operatorname{MAX}}^{(2)},\gamma}$.

\section{Simulation}
\subsection{One-sample problem}

For the one-sample problem, we compare our Fisher's combination test $T_{\textrm{FC}}^{(1)}$ in (\ref{DEF_T_FC}) (abbreviated as FC) with
\begin{itemize}
\item the sum-type test $T_{SR}$ in (\ref{ts1}) proposed by \cite{srivastava2009test} (abbreviated as SR);
\item the max-type tests $M_{\I_p}$, $M_{\hat{\O}^{1/2}}$ and $M_{\hat{\O}}$ based on (\ref{DEF_max}) (abbreviated as MAX1, MAX2 and MAX3, respectively);
\item the higher criticism test $T_{HC}$ by \cite{zhong2013tests} (abbreviated as HC):
\begin{align}\lbl{qiche}
T_{HC2}=\max_{s\in \mathcal{S}}\frac{T_{2n}(s)-\hat{\mu}(s)}{\hat{\sigma}(s)},
\end{align}
where $\mathcal{S}$ is a subset of the interval $(0,1)$,
\begin{align*}
T_{2n}(s)&=\sum_{j=1}^pn\bigg(\bar{\X}_j/\sigma_j\bigg)^2I \bigg(|\bar{\X}_j|\ge \sigma_j\sqrt{\lambda_s/n}\bigg),\\
\hat{\mu}(s)&=p\bigg\{2\lambda_p^{1/2}(s)\phi(\lambda_p^{1/2}(s))+2\bar{\Phi}(\lambda_p^{1/2}(s))\bigg\},\\
\hat{\sigma}^2(s)&=p\bigg\{2\bigg[\lambda_p^{3/2}(s)+3\lambda_p^{1/2}(s)\bigg]\phi(\lambda_p^{1/2}(s))+6\bar{\Phi}(\lambda_p^{1/2}(s))\bigg\}.
\end{align*}
Here $\lambda_s(p)=2s\log p$, and $\phi(\cdot)$, $\bar{\Phi}(\cdot)$ are the density and survival functions of the standard normal distribution, respectively.

\item the power enhancement test $J$ by \cite{fan2015power} (abbreviated as PE):
 \bea\lbl{keyia}
J=J_0+J_1,
\eea
 where the power enhancement component $J_0$ is
$
J_0=\sqrt{p}\sum_{j=1}^p\bar{\X}_j^2\hat{\sigma}_j^{-2}I (|\bar{\X}_j|>\hat{\sigma}_j\delta_{p,n}),
$
and $J_1$ is the standard Wald statistic
$
J_1=\frac{\bar{\X}^T \widehat{\var}^{-1}(\hat{\X})\bar{\X}-p}{2\sqrt{p}}.
$
Here $\hat{\sigma}_j^2$ is the sample variance of the $j$th coordinate of the population vector, $\delta_{p,n}$ is a thresholding parameter and $\widehat{\var}^{-1}(\hat{\X})$ is a consistent estimator of the asymptotic inverse covariance matrix of $\bar{\X}$.
\end{itemize}

The specific models for the covariance structure are following the settings in \cite{tony2014two}. For convenience, we collected them as follows. Let $ {D}=\bigg(d_{i, j}\bigg)$ be a diagonal matrix with diagonal elements $d_{i, i}=\operatorname{Unif}(1,3)$ for $i=1, \ldots, p .$ Denote by $\lambda_{\min }( {A})$ the minimum eigenvalue of a symmetric matrix $ {A}$.

\begin{itemize}
\item[(a)] Model 1 (block diagonal $\Omega$): $ {\Sigma}=\bigg(\sigma_{i, j}\bigg)$ where $\sigma_{i, i}=1, \sigma_{i, j}=0.8$ for $2(k-1)+1 \leqslant i \neq j \leqslant 2 k$ where $k=1, \ldots,[p / 2]$ and $\sigma_{i, j}=0$ otherwise.
\item[(b)] Model 2 ('bandable' $ {\Sigma}$): $ {\Sigma}=\bigg(\sigma_{i, j}\bigg)$ where $\sigma_{i, j}=0.6^{|i-j|}$ for $1 \leqslant i, j \leqslant p$.
\item[(c)] Model 3 (banded $\Omega): \Omega=\bigg(\omega_{i, j}\bigg)$ where $\omega_{i, i}=2$ for $i=1, \ldots, p, \omega_{i, i+1}=0.8$ for $i=1, \ldots, p-$ $1, \omega_{i, i+2}=0.4$ for $i=1, \ldots, p-2, \omega_{i, i+3}=0.4$ for $i=1, \ldots, p-3, \omega_{i, i+4}=0.2$ for $i=$ $1, \ldots, p-4, \omega_{i, j}=\omega_{j, i}$ for $i, j=1, \ldots, p$ and $\omega_{i, j}=0$ otherwise.
\item[(d)] Model 4 (sparse $ {\Sigma}$): $ {\Omega}=\bigg(\omega_{i, j}\bigg)$ where $\omega_{i, j}=0.6^{|i-j|}$ for $1 \leqslant i, j \leqslant p .  {\Sigma}= {D}^{1 / 2}  {\Omega}^{-1}  {D}^{1 / 2}$.
\item[(e)] Model 5 (sparse $ {\Sigma}$): $ {\Omega}^{1 / 2}=\bigg(a_{i, j}\bigg)$ where $a_{i, i}=1, a_{i, j}=0.8$ for $2(k-1)+1 \leqslant i \neq j \leqslant 2 k$, where $k=1, \ldots,[p / 2]$, and $a_{i, j}=0$ otherwise. $ {\Omega}= {D}^{1 / 2}  {\Omega}^{1 / 2}  {\Omega}^{1 / 2}  {D}^{1 / 2}$ and $ {\Sigma}= {\Omega}^{-1}$
\item[(f)] Model 6 (non-sparse case): $ {\Sigma}^{*}=\bigg(\sigma_{i, j}^{*}\bigg)$ where $\sigma_{i, i}^{*}=1, \sigma_{i, j}^{*}=0.8$ for $2(k-1)+1 \leqslant i \neq$ $j \leqslant 2 k$, where $k=1, \ldots,[p / 2]$, and $\sigma_{i, j}^{*}=0$ otherwise. $ {\Sigma}= {D}^{1 / 2}  {\Sigma}^{*}  {D}^{1 / 2}+ {E}+\delta  {I}$ with $\delta=$ $\left|\lambda_{\min }\bigg( {D}^{1 / 2}  {\Sigma}^{*}  {D}^{1 / 2}+ {E}\bigg)\right|+0.05$, where $ {E}$ is a symmetric matrix with the support of the off-diagonal entries chosen independently according to the Bernoulli(0.3) distribution with the values of the non-zero entries drawn randomly from Unif $(-0.2,0.2)$.
\item[(g)] Model 7 (non-sparse case): $ {\Sigma}^{*}=\bigg(\sigma_{i, j}^{*}\bigg)$ where $\sigma_{i, i}^{*}=1$ and $\sigma_{i, j}^{*}=|i-j|^{-5} / 2$ for $i \neq j$ $ {\Sigma}= {D}^{1 / 2}  {\Sigma}^{*}  {D}^{1 / 2}$
\item[(h)] Model 8 (non-sparse case): $ {\Sigma}= {D}^{1 / 2}\bigg( {F}+ {u}_{1} {u}_{1}^{\prime}+ {u}_{2} {u}_{2}^{\prime}+ {u}_{3}  {u}_{3}^{\prime}\bigg)  {D}^{1 / 2}$, where $ {F}=\bigg(f_{i, j}\bigg)$ is a $p \times p$ matrix with $f_{i, i}=1, f_{i, i+1}=f_{i+1, i}=0.5$ and $f_{i, j}=0$ otherwise, and $ {u}_{i}$ are orthonormal vectors for $i=1,2,3$.
\end{itemize}

For the generation of errors $ \bmv_{i} = (\varepsilon_{i1},\cdots,\varepsilon_{ip})^\top$, we consider three settings of $\varepsilon_{ij}$'s:
\begin{itemize}
\item[(1)] Normal distribution: $\varepsilon_{ij}\overset{i.i.d}{\sim}
N(0,1)$;
\item[(2)] standardized $t_5$ distribution: $\varepsilon_{ij}\overset{i.i.d}{\sim}
t(5)/\sqrt{5/3}$
\item[(3)] standardized mixture normal distribution: $\varepsilon_{ij}\overset{i.i.d}{\sim}
\{0.9N(0,1)+0.1N(0,9)\}/\sqrt{1.8}$
\end{itemize}

Table \ref{t1}-\ref{t2} report the empirical sizes of these tests with $n=120, p=100,200,300$. We found that SR, $M_{\I_p}$, $M_{\hat{\O}^{1/2}}$, $M_{\hat{\O}}$ and FC can control the empirical sizes very well. The empirical sizes of HC test are a little smaller than the nominal level. However, PE test can not control the empirical sizes in most case. So we do not compare it in the alternative hypothesis.

\begin{figure}
\centering
\includegraphics[width=5in,angle=0]{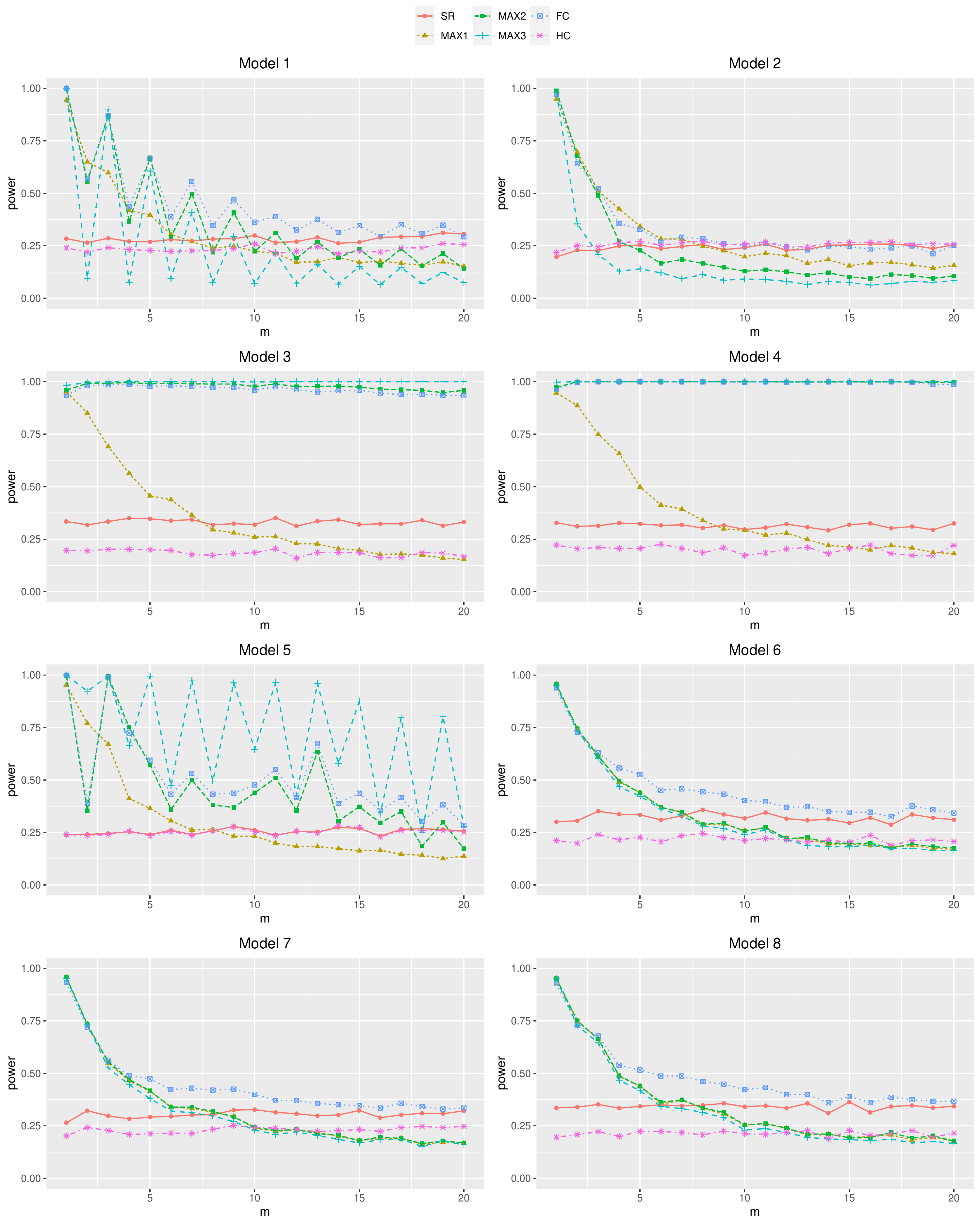}
\caption{Power of tests with different numbers of nonzero alpha at $n=120,p=200$ with normal errors. (MAX1 means $M_{\I_p}$; MAX2 means $M_{\hat{\O}^{1/2}}$; MAX3 means $M_{\hat{\O}}$.)}
\label{fig1}
\end{figure}

\begin{figure}
\centering
\includegraphics[width=5in,angle=0]{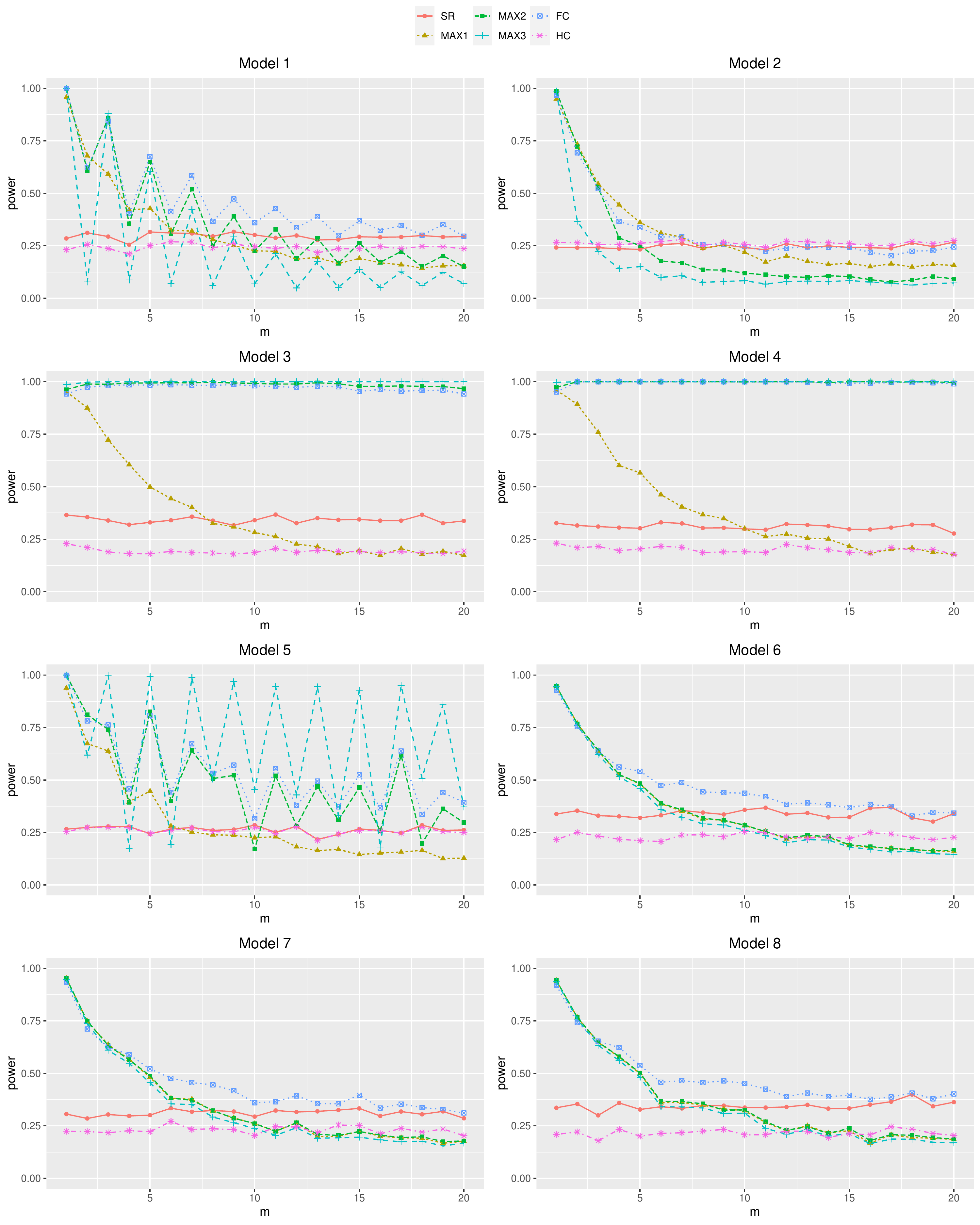}
\caption{Power of tests with different numbers of nonzero alpha at $n=120,p=200$ with $t(5)$ errors.(MAX1 means $M_{\I_p}$; MAX2 means $M_{\hat{\O}^{1/2}}$; MAX3 means $M_{\hat{\O}}$.)}
\label{fig2}
\end{figure}

\begin{figure}
\centering
\includegraphics[width=5in,angle=0]{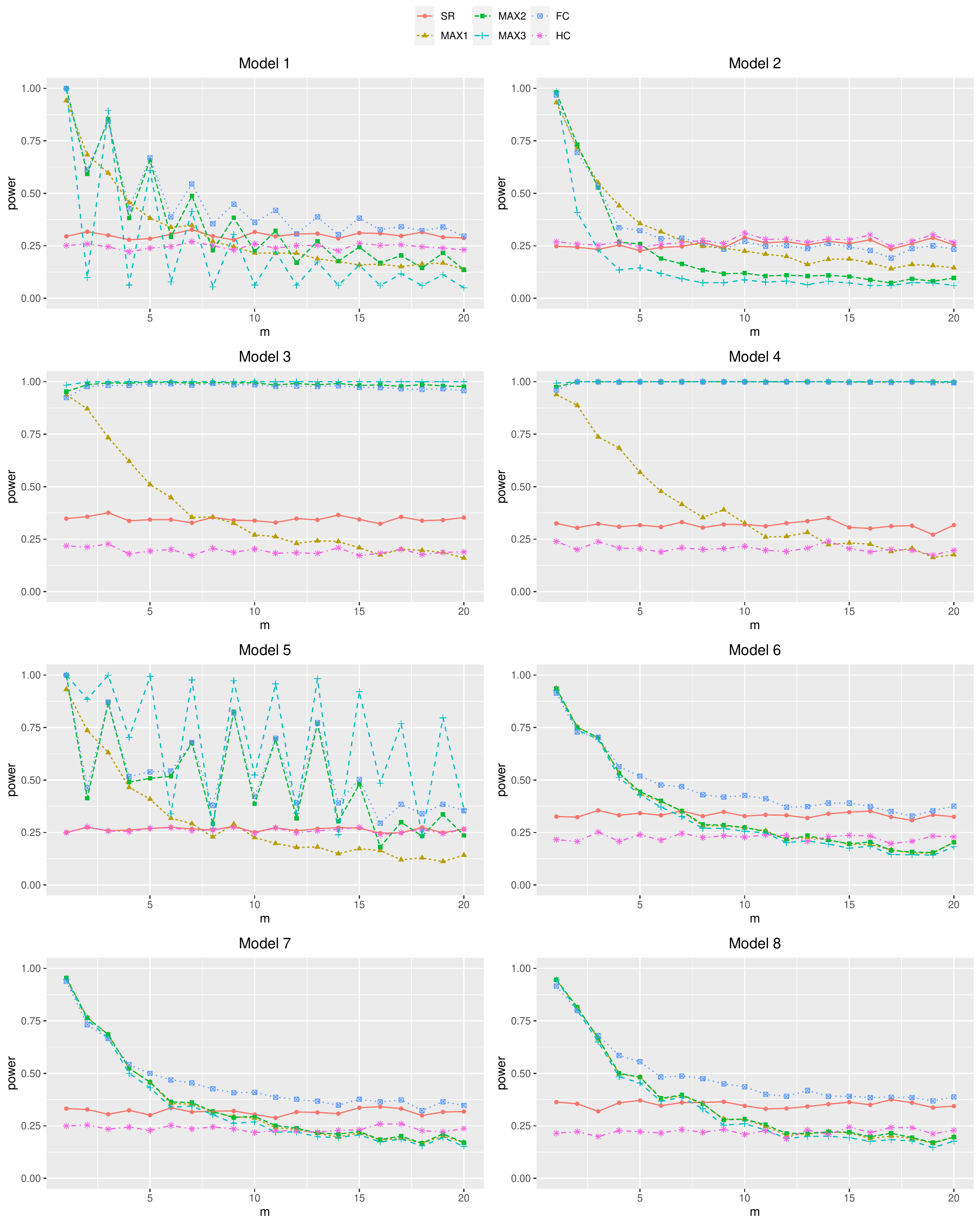}
\caption{Power of tests with different numbers of nonzero alpha at $n=120,p=200$ with mixture normal errors.(MAX1 means $M_{\I_p}$; MAX2 means $M_{\hat{\O}^{1/2}}$; MAX3 means $M_{\hat{\O}}$.)}
\label{fig3}
\end{figure}

For power comparison, we consider $ \bmu=\kappa(1/\sigma^{1/2}_{11},\cdots,1/\sigma^{1/2}_{mm},0,\cdots,0)$ where $\kappa$ is chosen as $||\bmu||^2=0.5$. Figures \ref{fig1}-\ref{fig3} report the power curves of each test with $n=120,p=200$ for different settings of the covariance structure. For the settings of error distribution, Figures \ref{fig1}, \ref{fig2} and \ref{fig3} report the power curves for the normal distribution, $t(5)$ distribution and mixture normal distribution, respectively. In general, the powers of the SR and HC tests are always very close to 0.25 under different choices of $m$ (ranging from 1 to 20). In contrast, under the most settings of the covariance structure, the powers of $M_{\I_p}$, $M_{\hat{\O}^{1/2}}$, $M_{\hat{\O}}$ and FC tests tends to decrease as $m$ increasing from 1 to 20 (except that under the Models 3 and 4, the powers of $M_{\hat{\O}^{1/2}}$, $M_{\hat{\O}}$ and FC tests are very close to 1 for difference choices of $m$). It is natural because the max-type tests can work better for the sparse case than the non-sparse case. In most scenarios, our proposed FC test is as powerful as the max-type tests when the number of variables with nonzero means is small and is more powerful than the max-type tests when the number of variables with nonzero means is large. This indicates that our FC test can work well in any case, which implies that our FC test is robust to the real data because it is not possible to tell if a dataset is sparse or not.

\begin{figure}
\centering
\includegraphics[width=5in,angle=0]{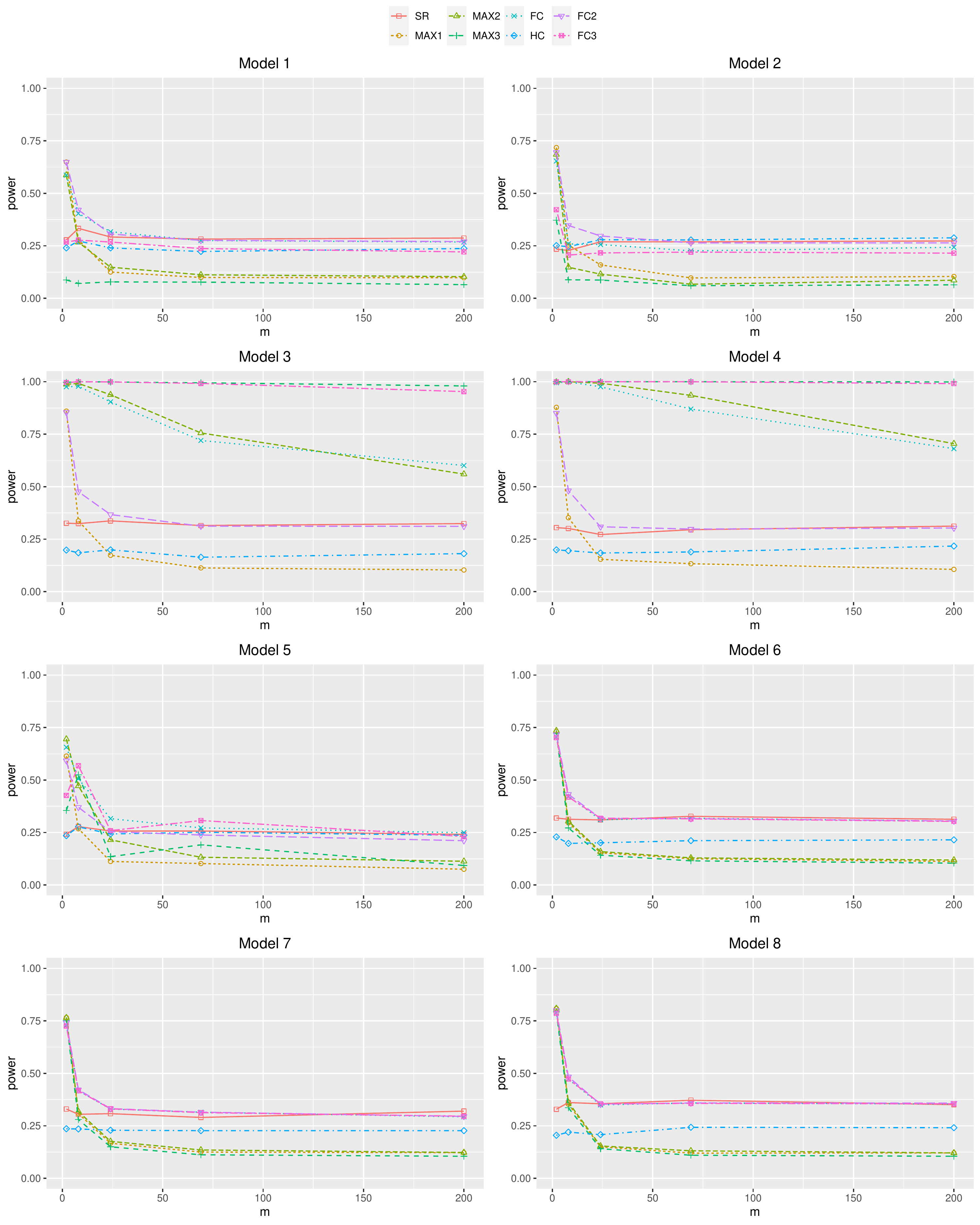}
\caption{Power of tests with different numbers of nonzero alpha at $n=120,p=200$ with normal errors and signal magnitude $||\mu||^2=0.5$.(MAX1 means $M_{\I_p}$; MAX2 means $M_{\hat{\O}^{1/2}}$; MAX3 means $M_{\hat{\O}}$.)}
\label{f22}
\end{figure}

\begin{figure}
\centering
\includegraphics[width=5in,angle=0]{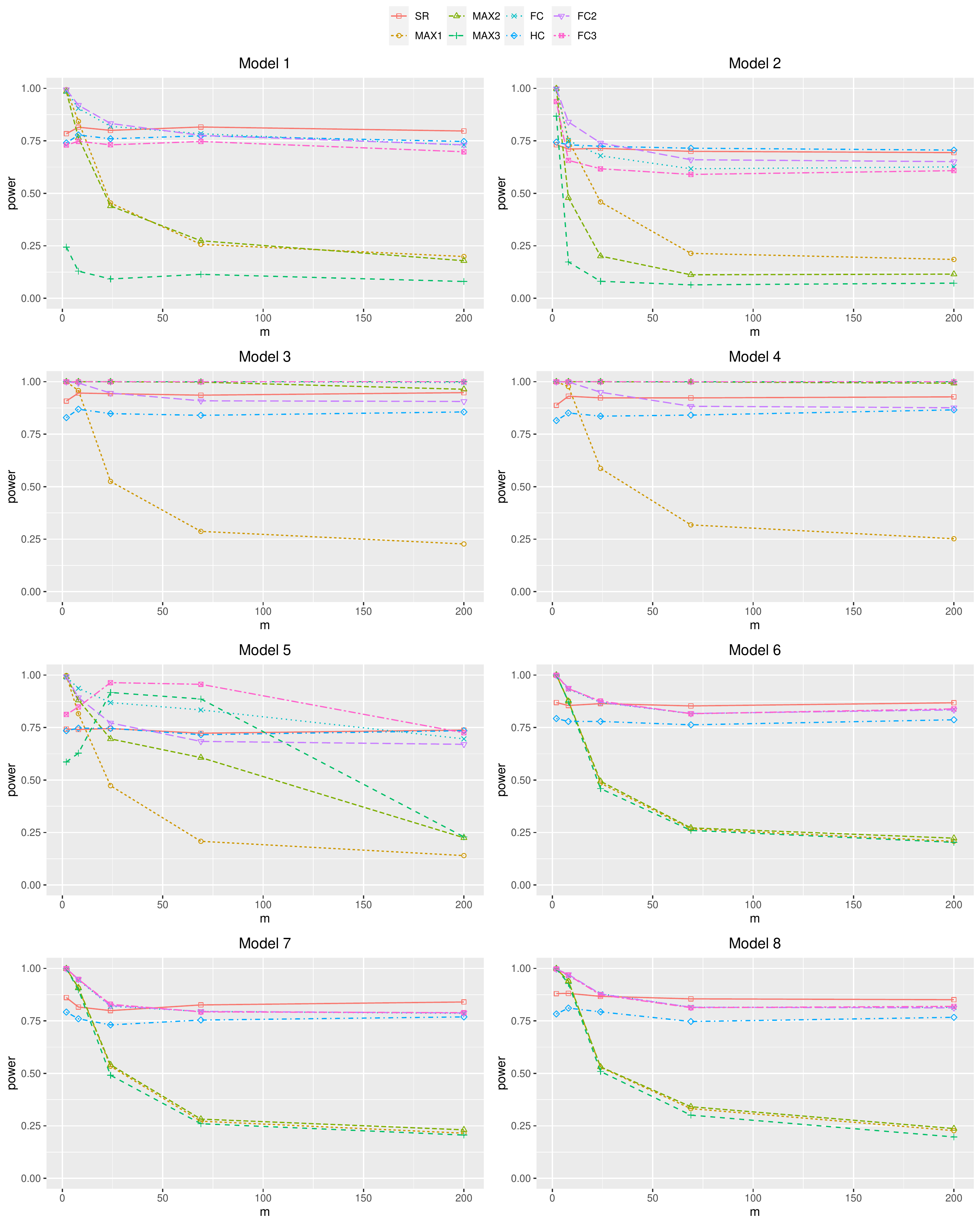}
\caption{Power of tests with different numbers of nonzero alpha at $n=120,p=200$ with normal errors and signal magnitude $||\mu||^2=0.8$.(MAX1 means $M_{\I_p}$; MAX2 means $M_{\hat{\O}^{1/2}}$; MAX3 means $M_{\hat{\O}}$.)}
\label{f23}
\end{figure}

In addition, we also consider the following two alternative Fisher's combination tests: (i) $T_{FC2}^{(1)}$ which is based on $T_{SR}$ and $M_{I_p}$ (abbreviated as FC2); (ii) $T_{FC3}^{(1)}$ which is based on $T_{SR}$ and $M_{\hat{\O}}$ (abbreviated as FC3). Table \ref{t22} reports the empirical sizes of FC2 and FC3 tests. We found that they both can control the empirical sizes in most cases. Additionally, Figures \ref{f22}, \ref{f23} report the power of eight tests with different numbers of nonzero alpha at $n=120,p=200$ with normal errors and different signal magnitude $||\mu||^2=0.5,0.8$, respectively. Now, we only consider $m=[p^a]$ where $a=1$ for the dense alternative,  $a=0.8,0.6$ for the median dense alternative , and $a=0.4,0.2$ for the sparse alternative. From Figure \ref{f22} and \ref{f23}, we found that the Fisher combination tests--FC,FC2 and FC3, perform better whether the alternative is dense or sparse.

\newpage
\subsection{Two-sample problem}

We compare our Fisher's combination test $T_{\textrm{FC}}^{(2)}$ in (\ref{DEF_T_FC_2}) (abbreviated as FC) with
\begin{itemize}
\item the sum-type test $T_{SKK}$ in (\ref{DEF_T_SKK}) proposed by \cite{srivastava2013two} (abbreviated as SKK);
\item the max-type tests $M_{\I_p}$, $M_{\hat{\O}^{1/2}}$ and $M_{\hat{\O}}$ proposed by \cite{tony2014two} (abbreviated as MAX1, MAX2 and MAX3, respectively);
\item the higher criticism test $T_{HC}$ by \cite{chen2019two} (abbreviated as HC);
\item the adaptive test $T_{AD}$ by \cite{xu2016adaptive} (abbreviated as AD).
\end{itemize}

We generate $\X_i= \bmu+ \bms^{1/2} \z_i$ and $\Y_i= \bms^{1/2}  \xi_i$ where $ \bms$ also generated from the eight models and $\z_i ,  \xi_i$ has the three scenarios as $ \bmv_i$ in the above subsection. Under the null hypothesis, we set $ \bmu=0$. Under the alternative hypothesis, we set $ \bmu= \bms^{1/2}  \theta$ where $  \theta=(\theta_1/\sqrt{m},\cdots,\theta_m/\sqrt{m},0,\cdots,0)$, $\theta_i \sim 2B(1,0.5)-1$ are independent binomial random variables.

Table \ref{t3}-\ref{t4} report the empirical sizes of these tests with $n_1=n_2=60, p=100,200,300$. The SKK, $M_{\hat{\O}^{1/2}}$, FC and AD tests can control the empirical sizes very well in most settings of the covariance structure. However, we also observe that (i) the empirical sizes of HC are a little smaller than the nominal level under the Models 1, 3, 4, 6, 7 and 8 of the covariance structure; (ii) the empirical sizes of $M_{\I_p}$ are a little smaller than the nominal level under the Models 1 and 5; (iii) the empirical sizes of $M_{\hat{\O}}$ are a little higher than the nominal level under the Models 2, 3 an 4.

\begin{figure}
\centering
\includegraphics[width=5in,angle=0]{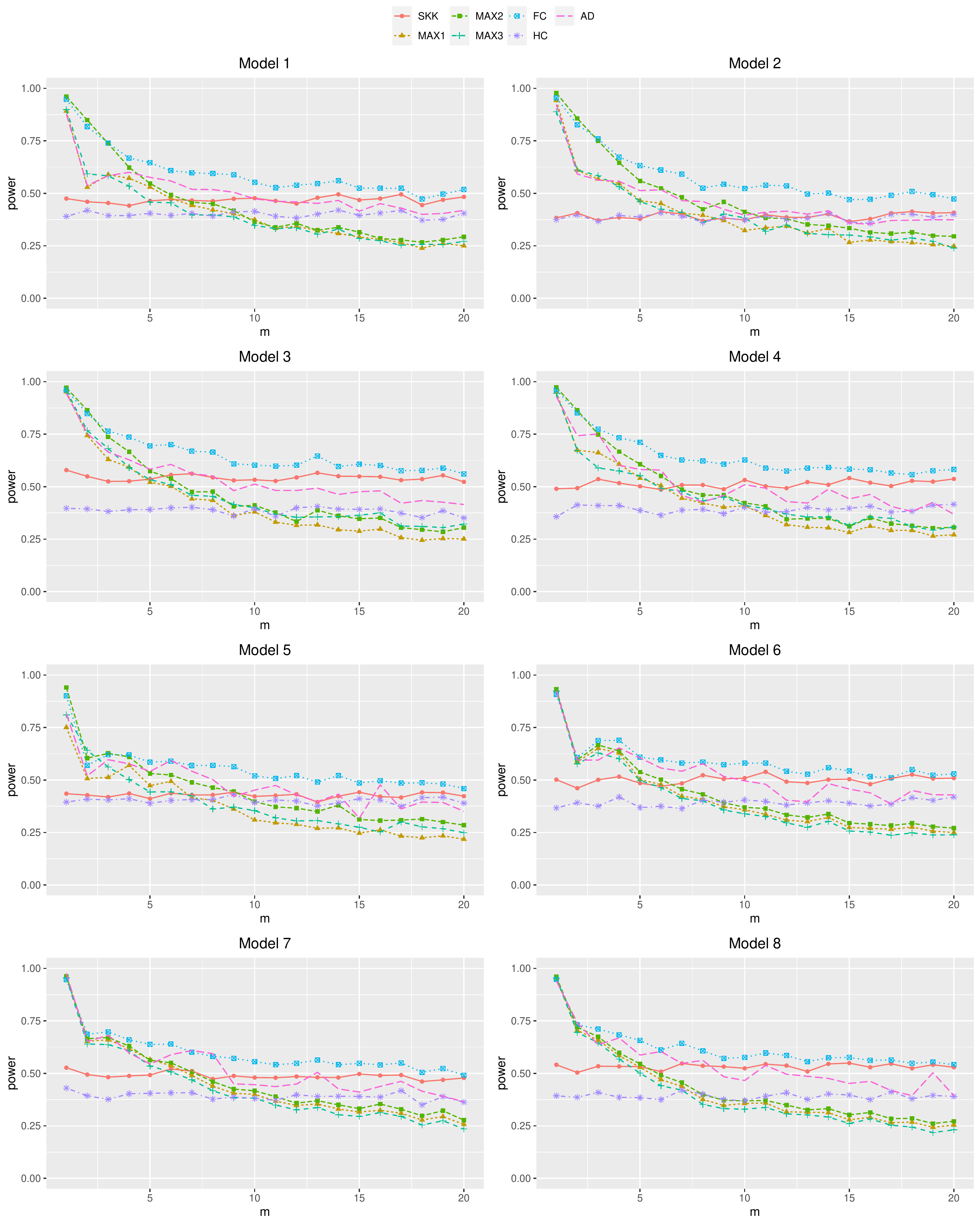}
\caption{Power of tests with different numbers of nonzero alpha at $n_1=n_2=60,p=100$ with normal errors.(MAX1 means $M_{\I_p}$; MAX2 means $M_{\hat{\O}^{1/2}}$; MAX3 means $M_{\hat{\O}}$.)}
\label{fig4}
\end{figure}

\begin{figure}
\centering
\includegraphics[width=5in,angle=0]{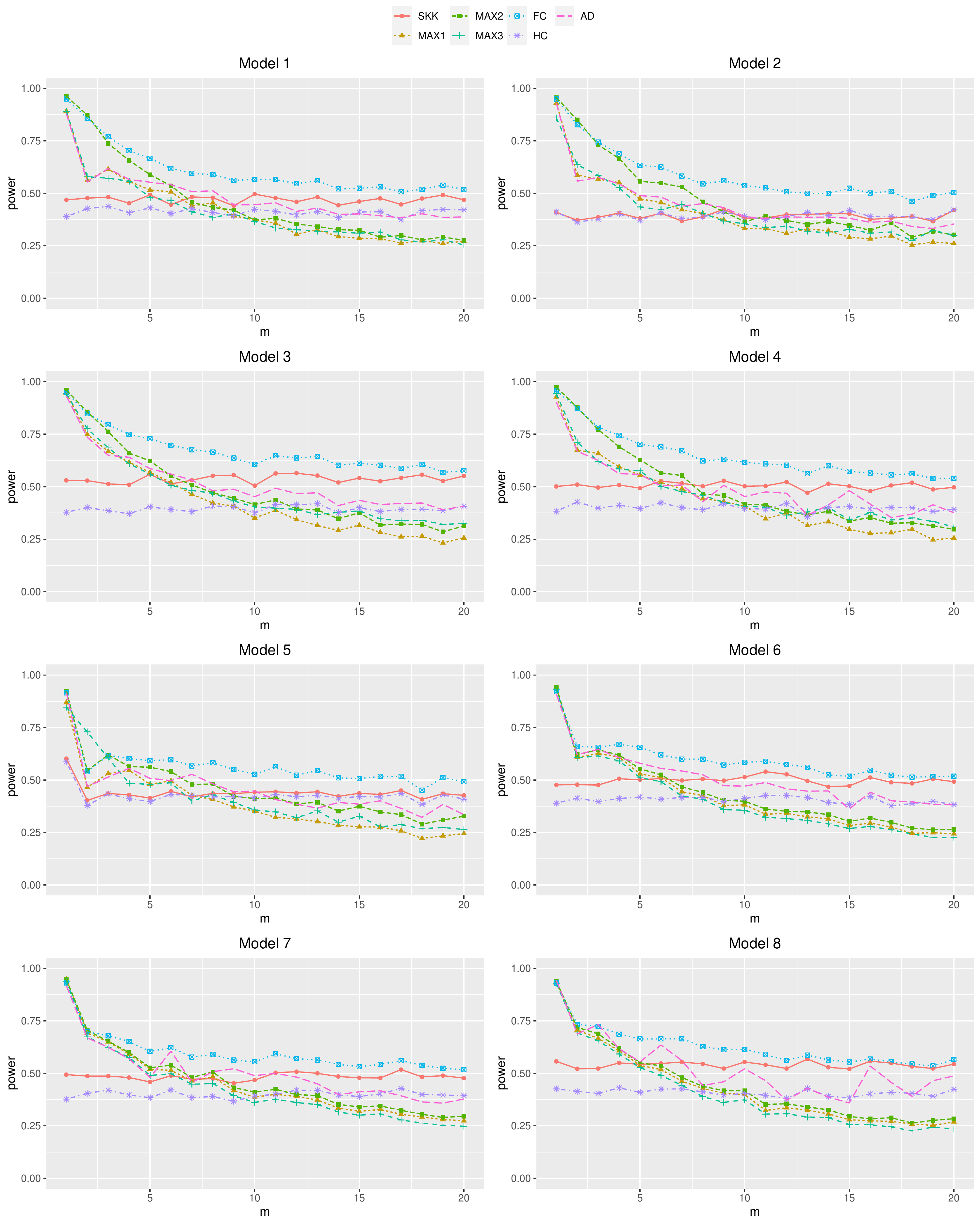}
\caption{Power of tests with different numbers of nonzero alpha at $n_1=n_2=60,p=100$ with $t(5)$ errors.(MAX1 means $M_{\I_p}$; MAX2 means $M_{\hat{\O}^{1/2}}$; MAX3 means $M_{\hat{\O}}$.)}
\label{fig5}
\end{figure}

\begin{figure}
\centering
\includegraphics[width=5in,angle=0]{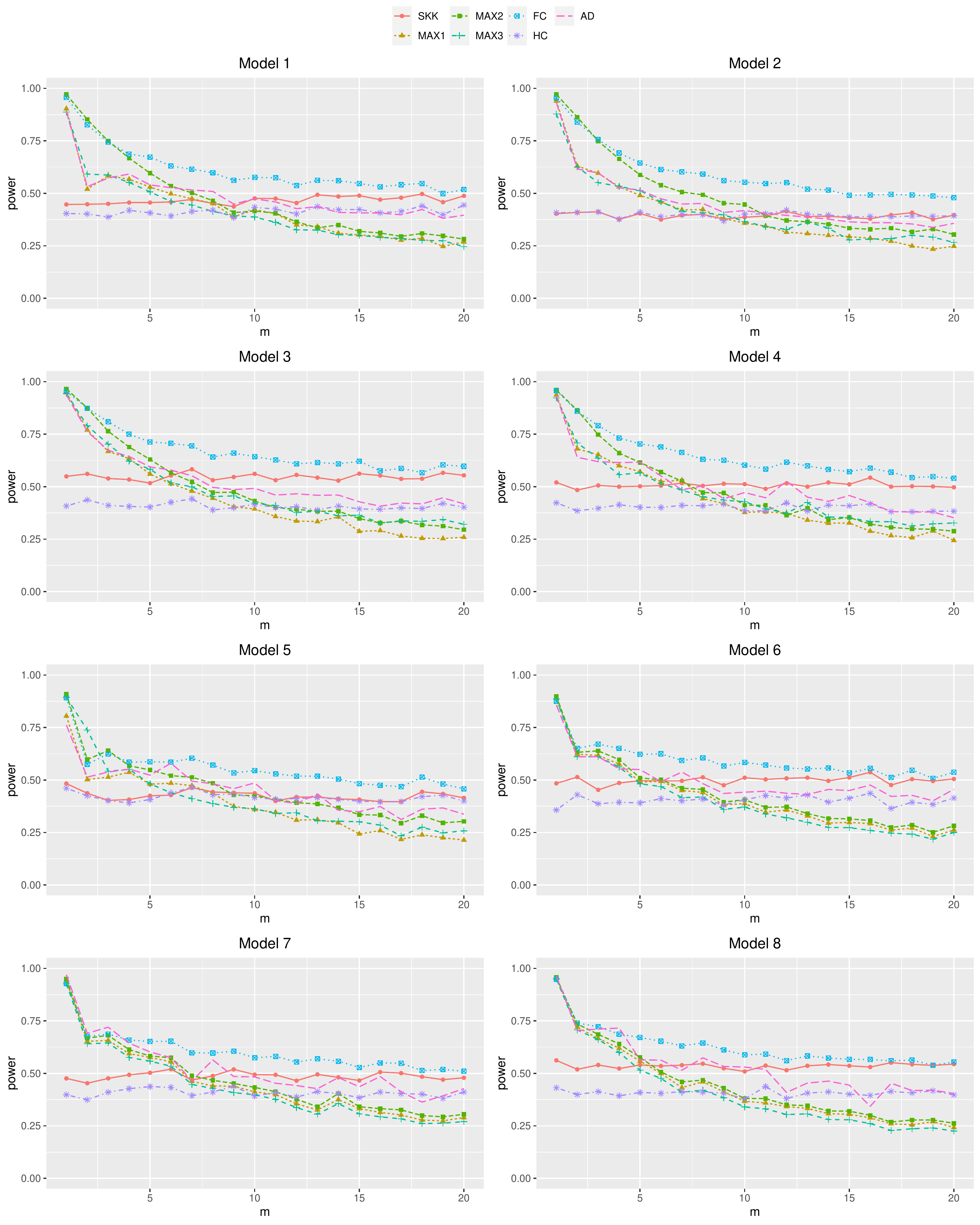}
\caption{Power of tests with different numbers of nonzero alpha at $n_1=n_2=60,p=100$ with mixture normal errors.(MAX1 means $M_{\I_p}$; MAX2 means $M_{\hat{\O}^{1/2}}$; MAX3 means $M_{\hat{\O}}$.)}
\label{fig6}
\end{figure}

Figures \ref{fig4}-\ref{fig6} reports the power curves of each test with $n_1=n_2=60, p=100$ for different settings of the covariance structure. For the settings of error distribution, Figures \ref{fig4}, \ref{fig5} and \ref{fig6} report the power curves for the normal distribution, $t(5)$ distribution and mixture normal distribution, respectively. In general, the powers of the SKK and HC tests are always staying around 0.5 under different choices of $m$ (ranging from 1 to 20) while the powers of other tests tend to decrease as $m$ increasing from 1 to 20. When the number of nonzero elements in $\tilde{  \mu}=  \theta$ is small, our proposed FC test is as powerful as the max-type tests. When the number of nonzero elements in $\tilde{  \mu}=  \theta$ is large, the power of our FC test is exactly higher than that of all other tests. In real world, it is impossible to identify whether the data is sparse or not. Thus, the above results demonstrate that our FC test is good in any case.

\begin{table}
\begin{center}

{
\begin{tabular}{c|ccc|ccc|ccc}
\hline \hline
  & \multicolumn{3}{c}{{Error (1)}} & \multicolumn{3}{c}{{Error (2)}}& \multicolumn{3}{c}{{Error (3)}} \\ \hline
$p$ &100&200&300&100&200&300&100&200&300\\
\hline
&\multicolumn{9}{c}{Model 1}\\ \hline
SR&0.049&0.038&0.04&0.044&0.041&0.041&0.049&0.056&0.042\\
$M_{\I_p}$&0.048&0.05&0.053&0.035&0.033&0.05&0.031&0.038&0.04\\
$M_{\hat{\O}^{1/2}}$&0.047&0.062&0.066&0.034&0.046&0.063&0.038&0.044&0.054\\
$M_{\hat{\O}}$&0.047&0.051&0.056&0.045&0.049&0.044&0.039&0.04&0.049\\
FC&0.063&0.053&0.063&0.046&0.056&0.055&0.056&0.056&0.056\\
HC&0.034&0.026&0.032&0.03&0.029&0.031&0.036&0.039&0.028\\
PE&0.062&0.079&0.124&0.095&0.131&0.171&0.105&0.148&0.202\\
&\multicolumn{9}{c}{Model 2}\\ \hline
SR&0.048&0.045&0.046&0.05&0.054&0.043&0.042&0.042&0.032\\
$M_{\I_p}$&0.048&0.053&0.055&0.029&0.057&0.053&0.042&0.046&0.043\\
$M_{\hat{\O}^{1/2}}$&0.068&0.064&0.068&0.058&0.044&0.061&0.043&0.046&0.059\\
$M_{\hat{\O}}$&0.06&0.06&0.077&0.051&0.048&0.076&0.049&0.056&0.071\\
FC&0.071&0.059&0.059&0.055&0.052&0.056&0.037&0.051&0.046\\
HC&0.055&0.048&0.048&0.055&0.058&0.048&0.045&0.05&0.038\\
PE&0.527&0.739&0.799&0.569&0.729&0.76&0.566&0.723&0.752\\
&\multicolumn{9}{c}{Model 3}\\ \hline
SR&0.046&0.051&0.055&0.046&0.045&0.04&0.043&0.031&0.034\\
$M_{\I_p}$&0.038&0.056&0.054&0.041&0.04&0.06&0.045&0.028&0.042\\
$M_{\hat{\O}^{1/2}}$&0.052&0.081&0.081&0.054&0.061&0.075&0.064&0.055&0.062\\
$M_{\hat{\O}}$&0.075&0.09&0.095&0.07&0.074&0.097&0.067&0.062&0.09\\
FC&0.062&0.077&0.089&0.06&0.061&0.06&0.07&0.048&0.056\\
HC&0.019&0.016&0.022&0.019&0.017&0.011&0.022&0.012&0.013\\
PE&0.048&0.088&0.126&0.051&0.082&0.118&0.063&0.086&0.113\\
&\multicolumn{9}{c}{Model 4}\\ \hline
SR&0.037&0.051&0.044&0.052&0.048&0.048&0.043&0.03&0.046\\
$M_{\I_p}$&0.04&0.053&0.065&0.045&0.048&0.05&0.035&0.038&0.046\\
$M_{\hat{\O}^{1/2}}$&0.049&0.077&0.09&0.063&0.055&0.057&0.042&0.05&0.045\\
$M_{\hat{\O}}$&0.056&0.077&0.071&0.057&0.065&0.075&0.057&0.061&0.064\\
FC&0.054&0.07&0.073&0.065&0.069&0.055&0.05&0.047&0.045\\
HC&0.019&0.033&0.021&0.034&0.02&0.022&0.023&0.009&0.024\\
PE&0.066&0.135&0.164&0.099&0.138&0.205&0.089&0.146&0.199\\
\hline
\hline
\end{tabular}}
\end{center}
\caption{\label{t1} Sizes of tests under model 1-4 in one-sample test.}
\end{table}

\begin{table}
\begin{center}
{
\begin{tabular}{c|ccc|ccc|ccc}
\hline \hline
  & \multicolumn{3}{c}{{Error (1)}} & \multicolumn{3}{c}{{Error (2)}}& \multicolumn{3}{c}{{Error (3)}} \\ \hline
$p$ &100&200&300&100&200&300&100&200&300\\
\hline
&\multicolumn{9}{c}{Model 5}\\ \hline
SR&0.052&0.044&0.035&0.047&0.043&0.039&0.049&0.033&0.039\\
$M_{\I_p}$&0.026&0.026&0.038&0.033&0.044&0.038&0.026&0.019&0.027\\
$M_{\hat{\O}^{1/2}}$&0.042&0.038&0.053&0.054&0.052&0.055&0.044&0.038&0.047\\
$M_{\hat{\O}}$&0.044&0.044&0.054&0.036&0.044&0.046&0.038&0.04&0.04\\
FC&0.056&0.05&0.052&0.052&0.063&0.055&0.057&0.047&0.054\\
HC&0.046&0.041&0.034&0.042&0.044&0.036&0.052&0.032&0.043\\
PE&0.012&0.001&0&0.023&0.004&0&0.021&0&0\\
&\multicolumn{9}{c}{Model 6}\\ \hline
SR&0.048&0.049&0.052&0.042&0.045&0.044&0.052&0.038&0.061\\
$M_{\I_p}$&0.048&0.046&0.051&0.034&0.042&0.057&0.038&0.041&0.054\\
$M_{\hat{\O}^{1/2}}$&0.056&0.05&0.06&0.034&0.043&0.059&0.041&0.044&0.056\\
$M_{\hat{\O}}$&0.045&0.042&0.057&0.03&0.033&0.053&0.034&0.037&0.052\\
FC&0.065&0.058&0.064&0.059&0.064&0.058&0.059&0.041&0.07\\
HC&0.026&0.02&0.024&0.025&0.027&0.022&0.027&0.014&0.028\\
PE&0.153&0.143&0.165&0.136&0.156&0.154&0.156&0.152&0.1\\
&\multicolumn{9}{c}{Model 7}\\ \hline
SR&0.052&0.048&0.048&0.046&0.045&0.029&0.056&0.06&0.048\\
$M_{\I_p}$&0.04&0.043&0.062&0.048&0.033&0.038&0.037&0.04&0.045\\
$M_{\hat{\O}^{1/2}}$&0.05&0.043&0.066&0.053&0.041&0.041&0.044&0.043&0.04\\
$M_{\hat{\O}}$&0.043&0.038&0.057&0.042&0.028&0.038&0.033&0.033&0.033\\
FC&0.054&0.048&0.06&0.053&0.053&0.038&0.053&0.051&0.049\\
HC&0.028&0.026&0.026&0.029&0.028&0.019&0.029&0.031&0.034\\
PE&0.152&0.163&0.185&0.148&0.153&0.156&0.185&0.186&0.179\\
&\multicolumn{9}{c}{Model 8}\\ \hline
SR&0.052&0.043&0.044&0.039&0.051&0.043&0.05&0.053&0.048\\
$M_{\I_p}$&0.037&0.047&0.056&0.042&0.049&0.047&0.03&0.037&0.026\\
$M_{\hat{\O}^{1/2}}$&0.044&0.047&0.058&0.042&0.058&0.053&0.034&0.049&0.029\\
$M_{\hat{\O}}$&0.035&0.036&0.051&0.035&0.046&0.042&0.024&0.037&0.026\\
FC&0.048&0.042&0.044&0.038&0.056&0.05&0.037&0.045&0.034\\
HC&0.023&0.013&0.015&0.015&0.02&0.017&0.025&0.021&0.013\\
PE&0.149&0.161&0.143&0.111&0.144&0.15&0.125&0.177&0.157\\
\hline
\hline
\end{tabular}}
\end{center}
\caption{\label{t2} Sizes of tests under model 5-8 in one-sample test.}
\end{table}

\begin{table}
\begin{center}
{
\begin{tabular}{c|ccc|ccc|ccc}
\hline \hline
  & \multicolumn{3}{c}{{Error (1)}} & \multicolumn{3}{c}{{Error (2)}}& \multicolumn{3}{c}{{Error (3)}} \\ \hline
$p$ &100&200&300&100&200&300&100&200&300\\
\hline
&\multicolumn{9}{c}{Model 1}\\ \hline
FC2&0.077&0.064&0.068&0.057&0.061&0.062&0.045&0.037&0.045\\
FC3&0.072&0.060&0.056&0.054&0.039&0.053&0.049&0.040&0.047\\\hline
&\multicolumn{9}{c}{Model 2}\\ \hline
FC2&0.075&0.064&0.071&0.064&0.079&0.074&0.056&0.073&0.053\\
FC3&0.063&0.054&0.053&0.043&0.064&0.074&0.041&0.056&0.046\\\hline
&\multicolumn{9}{c}{Model 3}\\ \hline
FC2&0.069&0.058&0.077&0.060&0.058&0.056&0.049&0.043&0.050\\
FC3&0.078&0.083&0.087&0.065&0.061&0.081&0.066&0.067&0.068\\\hline
&\multicolumn{9}{c}{Model 4}\\ \hline
FC2&0.066&0.078&0.073&0.059&0.055&0.069&0.060&0.053&0.050\\
FC3&0.060&0.065&0.073&0.059&0.065&0.069&0.046&0.067&0.057\\\hline
&\multicolumn{9}{c}{Model 5}\\ \hline
FC2&0.046&0.043&0.042&0.062&0.036&0.049&0.045&0.035&0.038\\
FC3&0.034&0.043&0.039&0.045&0.037&0.050&0.051&0.037&0.045\\\hline
&\multicolumn{9}{c}{Model 6}\\ \hline
FC2&0.070&0.067&0.06&0.073&0.046&0.063&0.048&0.056&0.056\\
FC3&0.072&0.06&0.056&0.071&0.044&0.058&0.048&0.055&0.056\\\hline
&\multicolumn{9}{c}{Model 7}\\ \hline
FC2&0.061&0.067&0.055&0.056&0.054&0.069&0.060&0.062&0.046\\
FC3&0.058&0.062&0.057&0.056&0.049&0.066&0.056&0.057&0.041\\\hline
&\multicolumn{9}{c}{Model 8}\\ \hline
FC2&0.058&0.070&0.054&0.064&0.062&0.061&0.071&0.044&0.050\\
FC3&0.058&0.062&0.052&0.058&0.067&0.058&0.065&0.043&0.052\\
\hline
\hline
\end{tabular}}
\end{center}
\caption{\label{t22} Sizes of FC2 and FC3 tests under models 1-8 in one-sample test.}
\end{table}

\begin{table}
\begin{center}
{
\begin{tabular}{c|ccc|ccc|ccc}
\hline \hline
  & \multicolumn{3}{c}{{Error (1)}} & \multicolumn{3}{c}{{Error (2)}}& \multicolumn{3}{c}{{Error (3)}} \\ \hline
$p$ &100&200&300&100&200&300&100&200&300\\
\hline
&\multicolumn{9}{c}{Model 1}\\ \hline
SKK&0.057&0.054&0.049&0.059&0.055&0.054&0.055&0.05&0.065\\
$M_{\I_p}$&0.047&0.058&0.051&0.031&0.033&0.037&0.035&0.034&0.031\\
$M_{\hat{\O}^{1/2}}$&0.057&0.063&0.065&0.052&0.059&0.063&0.051&0.055&0.047\\
$M_{\hat{\O}}$&0.037&0.063&0.072&0.041&0.052&0.054&0.046&0.051&0.039\\
FC&0.049&0.045&0.049&0.043&0.049&0.045&0.046&0.04&0.047\\
HC&0.046&0.035&0.055&0.043&0.035&0.03&0.042&0.035&0.025\\
AD&0.058&0.052&0.06&0.053&0.042&0.036&0.043&0.045&0.037\\
&\multicolumn{9}{c}{Model 2}\\ \hline
SKK&0.064&0.064&0.062&0.064&0.054&0.058&0.066&0.045&0.05\\
$M_{\I_p}$&0.04&0.054&0.059&0.054&0.045&0.04&0.04&0.047&0.051\\
$M_{\hat{\O}^{1/2}}$&0.073&0.081&0.089&0.066&0.082&0.078&0.058&0.062&0.067\\
$M_{\hat{\O}}$&0.047&0.067&0.076&0.041&0.053&0.072&0.05&0.054&0.071\\
FC&0.056&0.056&0.046&0.044&0.037&0.05&0.048&0.042&0.047\\
HC&0.066&0.064&0.054&0.063&0.049&0.065&0.053&0.058&0.055\\
AD&0.063&0.059&0.064&0.068&0.042&0.054&0.044&0.064&0.054\\
&\multicolumn{9}{c}{Model 3}\\ \hline
SKK&0.057&0.068&0.05&0.044&0.045&0.053&0.051&0.06&0.057\\
$M_{\I_p}$&0.042&0.053&0.054&0.042&0.056&0.04&0.047&0.038&0.034\\
$M_{\hat{\O}^{1/2}}$&0.06&0.088&0.103&0.066&0.068&0.061&0.058&0.065&0.067\\
$M_{\hat{\O}}$&0.054&0.101&0.11&0.073&0.079&0.081&0.059&0.076&0.078\\
FC&0.04&0.042&0.037&0.036&0.045&0.051&0.05&0.042&0.041\\
HC&0.024&0.034&0.022&0.023&0.015&0.019&0.027&0.028&0.023\\
AD&0.046&0.094&0.059&0.042&0.094&0.04&0.044&0.093&0.036\\
&\multicolumn{9}{c}{Model 4}\\ \hline
SKK&0.05&0.045&0.04&0.05&0.056&0.064&0.054&0.043&0.042\\
$M_{\I_p}$&0.033&0.049&0.049&0.045&0.04&0.044&0.037&0.047&0.041\\
$M_{\hat{\O}^{1/2}}$&0.065&0.079&0.082&0.061&0.075&0.072&0.053&0.071&0.065\\
$M_{\hat{\O}}$&0.084&0.071&0.086&0.05&0.08&0.066&0.061&0.074&0.078\\
FC&0.043&0.036&0.032&0.042&0.048&0.044&0.041&0.042&0.039\\
HC&0.027&0.024&0.025&0.044&0.027&0.028&0.033&0.03&0.019\\
AD&0.063&0.056&0.051&0.057&0.05&0.051&0.058&0.049&0.038\\
\hline
\hline
\end{tabular}}
\end{center}
\caption{\label{t3} Sizes of tests under model 1-4 in two-sample test.}
\end{table}

\begin{table}
\begin{center}
{
\begin{tabular}{c|ccc|ccc|ccc}
\hline \hline
  & \multicolumn{3}{c}{{Error (1)}} & \multicolumn{3}{c}{{Error (2)}}& \multicolumn{3}{c}{{Error (3)}} \\ \hline
$p$ &100&200&300&100&200&300&100&200&300\\
\hline
&\multicolumn{9}{c}{Model 5}\\ \hline
SKK&0.05&0.063&0.056&0.061&0.068&0.062&0.05&0.055&0.05\\
$M_{\I_p}$&0.03&0.036&0.045&0.029&0.028&0.033&0.026&0.023&0.027\\
$M_{\hat{\O}^{1/2}}$&0.046&0.069&0.079&0.05&0.067&0.052&0.054&0.052&0.049\\
$M_{\hat{\O}}$&0.052&0.084&0.07&0.044&0.064&0.057&0.047&0.043&0.053\\
FC&0.054&0.043&0.049&0.058&0.047&0.048&0.034&0.039&0.038\\
HC&0.04&0.057&0.062&0.045&0.046&0.056&0.056&0.051&0.052\\
AD&0.047&0.061&0.047&0.043&0.042&0.038&0.039&0.041&0.05\\
&\multicolumn{9}{c}{Model 6}\\ \hline
SKK&0.061&0.046&0.059&0.056&0.048&0.052&0.058&0.063&0.047\\
$M_{\I_p}$&0.038&0.047&0.056&0.046&0.038&0.051&0.035&0.037&0.041\\
$M_{\hat{\O}^{1/2}}$&0.041&0.055&0.063&0.052&0.045&0.058&0.042&0.046&0.05\\
$M_{\hat{\O}}$&0.033&0.042&0.052&0.04&0.036&0.045&0.034&0.034&0.035\\
FC&0.043&0.036&0.043&0.042&0.037&0.051&0.046&0.062&0.043\\
HC&0.03&0.027&0.025&0.022&0.019&0.03&0.028&0.027&0.023\\
AD&0.054&0.061&0.06&0.045&0.044&0.056&0.047&0.046&0.045\\
&\multicolumn{9}{c}{Model 7}\\ \hline
SKK&0.047&0.059&0.048&0.06&0.052&0.065&0.054&0.058&0.053\\
$M_{\I_p}$&0.049&0.047&0.062&0.049&0.044&0.049&0.039&0.04&0.044\\
$M_{\hat{\O}^{1/2}}$&0.053&0.052&0.069&0.053&0.05&0.055&0.042&0.046&0.05\\
$M_{\hat{\O}}$&0.039&0.038&0.055&0.045&0.04&0.047&0.031&0.036&0.045\\
FC&0.037&0.047&0.048&0.045&0.046&0.045&0.052&0.046&0.045\\
HC&0.028&0.031&0.026&0.031&0.03&0.028&0.033&0.035&0.022\\
AD&0.056&0.048&0.07&0.06&0.053&0.045&0.059&0.042&0.048\\
&\multicolumn{9}{c}{Model 8}\\ \hline
SKK&0.069&0.057&0.04&0.054&0.053&0.06&0.057&0.063&0.055\\
$M_{\I_p}$&0.052&0.054&0.051&0.043&0.044&0.042&0.037&0.036&0.04\\
$M_{\hat{\O}^{1/2}}$&0.063&0.058&0.056&0.047&0.048&0.048&0.041&0.038&0.046\\
$M_{\hat{\O}}$&0.049&0.05&0.041&0.035&0.042&0.033&0.033&0.03&0.038\\
FC&0.056&0.054&0.039&0.06&0.043&0.044&0.045&0.052&0.049\\
HC&0.025&0.024&0.03&0.032&0.027&0.023&0.016&0.021&0.019\\
AD&0.091&0.073&0.082&0.078&0.075&0.049&0.079&0.073&0.064\\
\hline
\hline
\end{tabular}}
\end{center}
\caption{\label{t4} Sizes of tests under model 5-8 in two-sample test.}
\end{table}

\newpage

\section{Appendix}
First, we restate the Central Limit Theorem for Linear Quadratic (\cite{kelejian2001asymptotic}, Theorem 1, p.227).
\begin{theorem}\label{th1}
Consider the following linear quadratic form
$$
Q_{p}=\varepsilon^{\prime}  {A} \varepsilon+ {b}^{\prime} \varepsilon=\sum_{i=1}^{p} \sum_{j=1}^{p} a_{i j} \varepsilon_{i} \varepsilon_{j}+\sum_{i=1}^{p} b_{i} \varepsilon_{i}
$$
where $\bigg\{\varepsilon_{i}, i=1,2, \ldots, p\bigg\}$ are real valued random variables, and $a_{i j}$ and $b_{i}$ denote real valued coefficients of the quadratic and linear forms. Suppose the following assumptions hold:  (i): $\varepsilon_{i}$, for $i=1,2, \ldots, p$, have zero means and are independently distributed across $i .$ (ii): A is symmetric and $\sup _{i} \sum_{j=1}^{p}\left|a_{i j}\right|<K .$ Also $p^{-1} \sum_{i=1}^{p}\left|b_{i}\right|^{2+\varepsilon_{0}}<K$ for some $\varepsilon_{0}>0 .$ (iii): $\sup _{i} E\left|\varepsilon_{i}\right|^{4+\varepsilon_{0}}<K$ for some $\varepsilon_{0}>0 .$ Then, assuming that $p^{-1} \operatorname{Var}\bigg(Q_{p}\bigg) \geq c$ for some $c>0$
$$
\frac{Q_{p}-E\bigg(Q_{p}\bigg)}{\sqrt{\operatorname{Var}\bigg(Q_{p}\bigg)}} \rightarrow_{d} N(0,1)
$$
\end{theorem}

\subsection{Proof of Theorem \ref{ind}}
Define $B_i=\{|z_i|>l_p\}$ and $A_p(x)=\bigg\{\frac{\z^\top\A\z-\tr(\A)}{\sigma_A}\le x\bigg\}$. We first prove the following important lemma.
\begin{lemma}\label{le1}
Under the assumption of Theorem \ref{ind}, for each $d\ge 1$, we have
\begin{align}\label{le11}
\lim_{p\to \infty} H(d,p) \le \frac{1}{d!} h^d(y)<\infty
\end{align}
where $H(d,p)\doteq \sum_{1\le i_1<\cdots<i_d\le p} P(B_{i_1}\cdots B_{i_d})$.
And then, we have
\begin{equation}\label{le12}
\sum_{1 \leq i_{1}<\cdots<i_{d} \leq p}\left|P\bigg(A_{p}(x) B_{i_{1}} \cdots B_{i_{d}}\bigg)-P\bigg(A_{p}(x)\bigg) \cdot P\bigg(B_{i_{1}} \cdots B_{i_{d}}\bigg)\right| \rightarrow 0
\end{equation}
as $p\to \infty$.
\end{lemma}
\proof Because $pP(B_i)\to h(y)$, we have $pP(B_i)<h(y)+\epsilon$ for any $\epsilon>0$ as $p\to \infty$. By the independence of $z_i$, we have
\begin{align*}
H(d,p)=&\sum_{1\le i_1<\cdots<i_d\le p} P(B_{i_1}\cdots B_{i_d})=\sum_{1\le i_1<\cdots<i_d\le p} \prod_{k=1}^d P(B_{i_k})\\
\le& C_p^d \{p^{-1}(h(y)+\epsilon)\}^d\le \frac{1}{d!} \bigg(h(y)+\epsilon\bigg)^d
\end{align*}
So, by letting $\epsilon \to 0$, we have
\begin{align*}
\lim_{p\to \infty}H(d,p) \le \frac{1}{d!} h^d(y)<\infty
\end{align*}
by assumption (i) in Theorem \ref{ind}. Here we prove (\ref{le11}).

Define $\z=(\z_1,\z_2)$ where $\z_1=(z_1,\cdots,z_d)$ and $\z_2=(z_{d+1},\cdots,z_p)$. And
\begin{align*}
\A=\bigg(
\begin{array}{cc}
\A_{11}&\A_{12}\\
\A_{21}&\A_{22}
\end{array}
\bigg)
\end{align*}
So,
\begin{align*}
\z^\top\A\z=\z_{1}^\top \A_1\z_1+2\z_1^\top \A_{12}\z_2+\z_2^\top \A_2\z_2.
\end{align*}
Next, we will show that
\begin{align*}
P\bigg(\z_{1}^\top \A_1\z_1>\epsilon\sigma_A\bigg)\le p^{-t}
\end{align*}
for $\epsilon>0$. Because $z_i$ is sub-Gaussian random variables, there exist $\eta>0$ and $K>0$ such that $E(\exp(\eta z^2_i))\le K$. Because $\lambda_{\max}(\A_1)\le \lambda_{\max}(\A)<c$,
\begin{align*}
P\bigg(\z_{1}^\top \A_1\z_1>\epsilon\sigma_A\bigg)\le& P\bigg(c\z_{1}^\top \z_1>\epsilon\sigma_A\bigg)\\
= & P\bigg(\eta\sum_{i=1}^d z_i^2>c^{-1}\eta\epsilon\sigma_A\bigg)\\
\le & \exp\bigg(-c^{-1}\eta\epsilon\sigma_A\bigg) E(e^{\eta\sum_{i=1}^d z_i^2})\\
=&\exp\bigg(-c^{-1}\eta\epsilon\sigma_A\bigg) \{E(e^{\eta z_i^2})\}^d\\
\le & K^{d}\exp\bigg(-c^{-1}\eta\epsilon\sigma_A\bigg)
\end{align*}
By the assumption (iii), we have $\sigma_A^2\ge 2\tr(\A^2)\ge 2c^{-2}p$. So
\begin{align}\label{s1}
P\bigg(\z_{1}^\top \A_1\z_1>\epsilon\sigma_A\bigg)\le K^{d}\exp\bigg(-\sqrt{2}c^{-2}\eta\epsilon p^{1/2}\bigg).
\end{align}
Define $ {A}= {Q}^{\top}  {\Lambda}  {Q}$ where $ {Q}=\bigg(q_{i j}\bigg)_{1 \leq i, j \leq p}$ is an orthogonal matrix and $ {\Lambda}=\operatorname{diag}\bigg\{\lambda_{1}, \ldots, \lambda_{p}\bigg\}, \lambda_{i}, i=1, \ldots, p$ are the eigenvalues of $ {A}$. Note that $\sum_{1 \leq j \leq p} a_{i j}^{2}$ is the $i$ th diagonal element of $ {A}^{2}= {Q}^{\top}  {\Lambda}^{2}  {Q}$, we have $\sum_{1 \leq j \leq p} a_{i j}^{2}=\sum_{l=1}^{p} q_{l i}^{2} \lambda_{l}^{2} \leq c^{2}$ according to Assumption (iii).

Next, define $\theta=\sqrt{\frac{2\eta}{dc^2\sigma^2}}$, we have
\begin{align*}
P\bigg(\z_1^\top \A_{12}\z_2\ge \epsilon\sigma_A\bigg)\le& \exp\bigg(-\theta\epsilon\sigma_A\bigg) E\bigg(\exp(\theta\z_1^\top \A_{12}\z_2)\bigg)\\
=&\exp\bigg(-\theta\epsilon\sigma_A\bigg) E(e^{\theta\sum_{i=1}^d\sum_{j=d+1}^pa_{ij}z_iz_j})\\
\le &\exp\bigg(-\theta\epsilon\sigma_A\bigg) E(E(e^{\theta\sum_{j=d+1}^p(\sum_{i=1}^da_{ij}z_i)z_j}|\z_j))\\
=&\exp\bigg(-\theta\epsilon\sigma_A\bigg) E\bigg(\prod_{j=d+1}^{p} E(e^{(\theta\sum_{i=1}^da_{ij}z_i)z_j}|\z_j)\bigg)\\
\le &\exp\bigg(-\theta\epsilon\sigma_A\bigg) E\bigg(\prod_{j=d+1}^{p} \exp\bigg(\frac{\sigma^2\theta^2}{2}\bigg(\sum_{i=1}^da_{ij}z_i\bigg)^2\bigg)\bigg)\\
=&\exp\bigg(-\theta\epsilon\sigma_A\bigg) E\bigg( \exp\bigg(\frac{\sigma^2\theta^2}{2}\sum_{j=d+1}^{p} \bigg(\sum_{i=1}^da_{ij}z_i\bigg)^2\bigg)\bigg)\\
\le &\exp\bigg(-\theta\epsilon\sigma_A\bigg) E\bigg( \exp\bigg(\frac{d\sigma^2\theta^2}{2}\sum_{j=d+1}^{p}\sum_{i=1}^da_{ij}^2z_i^2\bigg)\bigg)\\
\le &\exp\bigg(-\theta\epsilon\sigma_A\bigg) E\bigg( \exp\bigg(\frac{dc^2\sigma^2\theta^2}{2}\sum_{i=1}^dz_i^2\bigg)\bigg)\\
= &\exp\bigg(-\theta\epsilon\sigma_A\bigg) E\bigg( \exp\bigg(\eta\sum_{i=1}^dz_i^2\bigg)\bigg)\\
\le &K^d\exp\bigg(-\theta\epsilon\sigma_A\bigg)\le K^d \exp\bigg(-\sqrt{2}c^{-1}\theta\epsilon p^{1/2}\bigg)
\end{align*}
So
\begin{align}\label{s2}
P\bigg(\z_1^\top \A_{12}\z_2\ge \epsilon\sigma_A\bigg)\le&K^d \exp\bigg(-\sqrt{\frac{4\eta}{dc^4\sigma^2}}\epsilon p^{1/2}\bigg)
\end{align}
Similarly, we also can prove that
\begin{align}\label{s3}
P\bigg((-\z_1)^\top \A_{12}\z_2\ge \epsilon\sigma_A\bigg)\le&K^d \exp\bigg(-\sqrt{\frac{4\eta}{dc^4\sigma^2}}\epsilon p^{1/2}\bigg)
\end{align}
Let $\Theta_p=\z_{1}^\top \A_1\z_1+2\z_1^\top \A_{12}\z_2$.
\begin{align*}
P\bigg(|\Theta_p|>\epsilon\sigma_A\bigg)\le& P\bigg(\z_{1}^\top \A_1\z_1>\epsilon\sigma_A/2\bigg)+P\bigg(|\z_1^\top \A_{12}\z_2|>\epsilon\sigma_A/4\bigg)\\
\le &P\bigg(\z_{1}^\top \A_1\z_1>\epsilon\sigma_A/2\bigg)+P\bigg(\z_1^\top \A_{12}\z_2>\epsilon\sigma_A/8\bigg)+P\bigg(-\z_1^\top \A_{12}\z_2>\epsilon\sigma_A/8\bigg)\\
\end{align*}
So, by (\ref{s1}), (\ref{s2}) and (\ref{s3}), there exist a constant $c_{\epsilon}>0$,
\begin{align}\label{s4}
P\bigg(|\Theta_p|>\epsilon\sigma_A\bigg)\le K^d\exp(-c_{\epsilon}p^{1/2})
\end{align}
\begin{align*}
&P\bigg(A_{p}(x) B_{1} \cdots B_{d}\bigg)\\
=&P\bigg(\frac{\z_2^\top \A_2\z_2-\tr(\A)+\Theta_p}{\sigma_A}\le x, B_1\cdots B_d\bigg)\\
\le &P\bigg(\frac{\z_2^\top \A_2\z_2-\tr(\A)+\Theta_p}{\sigma_A}\le x, |\Theta_p|\le\epsilon \sigma_A, B_1\cdots B_d\bigg)+P\bigg(|\Theta_p|>\epsilon \sigma_A\bigg)\\
\le &P\bigg(\frac{\z_2^\top \A_2\z_2-\tr(\A)}{\sigma_A}\le x+\epsilon, B_1\cdots B_d\bigg)+K^d\exp(-c_{\epsilon}p^{1/2})\\
=&P\bigg(\frac{\z_2^\top \A_2\z_2-\tr(\A)}{\sigma_A}\le x+\epsilon\bigg)P\bigg(B_1\cdots B_d\bigg)+K^d\exp(-c_{\epsilon}p^{1/2})\\
\le &\bigg[P\bigg(\frac{\z_2^\top \A_2\z_2-\tr(\A)}{\sigma_A}\le x+\epsilon,|\Theta_p|\le\epsilon \sigma_A\bigg)+P\bigg(|\Theta_p|>\epsilon \sigma_A\bigg)\bigg]P\bigg(B_1\cdots B_d\bigg)\\
&+K^d\exp(-c_{\epsilon}p^{1/2})\\
\le & P\bigg(\frac{\z_2^\top \A_2\z_2-\tr(\A)+\Theta_p}{\sigma_A}\le x+2\epsilon\bigg)P\bigg(B_1\cdots B_d\bigg)+2K^d\exp(-c_{\epsilon}p^{1/2})\\
=&P\bigg(A_p( x+2\epsilon)\bigg)P\bigg(B_1\cdots B_d\bigg)+2K^d\exp(-c_{\epsilon}p^{1/2})
\end{align*}
Similarly, we can prove that
\begin{align*}
P\bigg(A_{p}(x) B_{1} \cdots B_{d}\bigg)\ge P\bigg(A_p( x-2\epsilon)\bigg)P\bigg(B_1\cdots B_d\bigg)-2K^d\exp(-c_{\epsilon}p^{1/2})
\end{align*}
So, we have
\begin{align}\label{pba}
\left|P\bigg(A_{p}(x) B_{1} \cdots B_{d}\bigg)-P\bigg(A_{p}(x)\bigg) \cdot P\bigg(B_{1} \cdots B_{d}\bigg)\right|
\leq  \Delta_{p, \epsilon} \cdot P\bigg(B_{1} \cdots B_{d}\bigg)+2K^d\exp(-c_{\epsilon}p^{1/2})
\end{align}
where
\begin{align*}
\Delta_{p, \epsilon}&=\left|P\bigg(A_{p}(x)\bigg)-P\bigg(A_{p}(x+2 \epsilon)\bigg)\right|+\left|P\bigg(A_{p}(x)\bigg)-P\bigg(A_{p}(x-2 \epsilon)\bigg)\right| \\
&=P\bigg(A_{p}(x+2 \epsilon)\bigg)-P\bigg(A_{p}(x-2 \epsilon)\bigg)
\end{align*}
Obviously, the inequality (\ref{pba}) holds for all $i_1,\cdots,i_d$. Thus,
\begin{align*}
&\sum_{1 \leq i_{1}<\cdots<i_{d} \leq p}\left|P\bigg(A_{p}(x) B_{i_{1}} \cdots B_{i_{d}}\bigg)-P\bigg(A_{p}(x)\bigg) \cdot P\bigg(B_{i_{1}} \cdots B_{i_{d}}\bigg)\right| \\
&\leq \sum_{1 \leq i_{1}<\cdots<i_{d} \leq p}\bigg[\Delta_{p, \epsilon} \cdot P\bigg(B_{i_{1}} \cdots B_{i_{d}}\bigg)+2K^d\exp(-c_{\epsilon}p^{1/2})\bigg] \\
&\leq \Delta_{p, \epsilon} \cdot H(d, p)+\bigg(\begin{array}{l}
p \\
d
\end{array}\bigg) \cdot 2K^d\exp(-c_{\epsilon}p^{1/2})
\end{align*}
By Theorem \ref{th1}, we have $P(A_p(x))\to \Phi(x)$ as $p\to \infty$. So $\Delta_{p, \epsilon}\to \Phi(x+2\epsilon)-\Phi(x-2\epsilon)$. By letting $\epsilon \to 0$, we have $\Delta_{p, \epsilon}\to 0$. By (\ref{le11}), we have $\lim_{p\to \infty}H(d,p)<\infty$. Additionally, $\bigg(\begin{array}{l}
p \\
d
\end{array}\bigg) \cdot 2K^d\exp(-c_{\epsilon}p^{1/2})\to 0$ as $p\to\infty$. So we can obtain (\ref{le12}).

{\it Proof of Theorem \ref{ind}}
First, we show that
\begin{align}\label{i2}
P\bigg(\max_{1\le i\le p}|z_i|\le l_p(y)\bigg)\to F(y).
\end{align}
Because $pP(|z_i|> l_p(y))\to h(y)$, we have $h(y)-\epsilon<pP(|z_i|> l_p(y))<h(y)+\epsilon$ for any $\epsilon>0$ as $p\to \infty$.
In fact, by the independence of $z_i$, we have
\begin{align*}
P\bigg(\max_{1\le i\le p}|z_i|\le l_p(y)\bigg)=&P\bigg(|z_i|\le l_p(y),1\le i\le p\bigg)=\prod_{i=1}^p\{P\bigg(|z_i|\le l_p(y)\bigg)\}\\
=&\prod_{i=1}^p(1-P\bigg(|z_i|> l_p(y)\bigg))\le (1-(h(y)-\epsilon)p^{-1})^p  \to e^{-h(y)+\epsilon}.
\end{align*}
Similarly, we have
\begin{align*}
P\bigg(\max_{1\le i\le p}|z_i|\le l_p(y)\bigg)=&\prod_{i=1}^p(1-P\bigg(|z_i|> l_p(y)\bigg))\ge (1-(h(y)+\epsilon)p^{-1})^p  \to e^{-h(y)-\epsilon}.
\end{align*}
So
\begin{align*}
e^{-h(y)-\epsilon}\le P\bigg(\max_{1\le i\le p}|z_i|\le l_p(y)\bigg)\le e^{-h(y)+\epsilon}.
\end{align*}
By letting $\epsilon\to 0$, we obtain the result (\ref{i2}).

 Additionally, by Theorem \ref{th1}, we know that
\begin{align}\label{i3}
P\bigg(\frac{\z^\top\A\z-\tr(\A)}{\sigma_A}\le x\bigg)\to \Phi(x)
\end{align}
To show (\ref{th11}), we only need to show that
\begin{align}\label{th12}
P\bigg(\frac{\z^\top\A\z-\tr(\A)}{\sigma_A}\le x,\max_{1\le i\le p}|z_i|> l_p(y)\bigg)\to\Phi(x)(1-F(y))
\end{align}
Recall the notations in Lemma \ref{le1}, we have
\bea\lbl{abci}
P\Big(\frac{\z^\top\A\z-\tr(\A)}{\sigma_A}\leq x,\ \max_{1\le i\le p}|z_i|>l_p\Big)=P\Big(\bigcup_{i=1}^pA_pB_{i}\Big).
\eea
Here the notation $A_pB_i$ stands for $A_p\cap B_i$ and we brief $A_p(x)$ as $A_p$.  From the inclusion-exclusion principle,
\bea
P\Big(\bigcup_{i=1}^pA_pB_{i}\Big)  \leq  \sum_{1\leq i_1 \leq p}P(A_pB_{i_1})&-&\sum_{1\leq i_1< i_2\leq p}P(A_pB_{i_1}B_{i_2})+\cdots+\nonumber\\
& & \sum_{1\leq i_1<  \cdots < i_{2k+1}\leq p}P(A_pB_{i_1}\cdots B_{i_{2k+1}})\nonumber\\
&&  \lbl{Upper_bound}
\eea
and
\bea
P\Big(\bigcup_{i=1}^pA_pB_{i}\Big)  \geq  \sum_{1\leq i_1 \leq p}P(A_pB_{i_1})&-&\sum_{1\leq i_1< i_2\leq p}P(A_pB_{i_1}B_{i_2})+\cdots- \nonumber\\
& & \sum_{1\leq i_1<  \cdots < i_{2k}\leq  p}P(A_pB_{i_1}\cdots B_{i_{2k}}) \nonumber\\
&&  \lbl{Lower_bound}
\eea
for any integer $k\geq 1$. Define
\beaa
H(p, d)=\sum_{1\leq i_1<  \cdots < i_{d}\leq p}P(B_{i_1}\cdots B_{i_{d}})
\eeaa
for $d\geq 1$. From \eqref{le11} we know
%
%
\bea\lbl{Maya1}
\lim_{d\to\infty}\limsup_{p\to\infty}H(p, d)=0.
\eea
Set
\beaa
\zeta(p,d)=\sum_{1\leq i_1<  \cdots < i_d\leq p}\big[P(A_pB_{i_1}\cdots B_{i_d}) - P(A_p)\cdot P(B_{i_1}\cdots B_{i_d})\big]
\eeaa
for $d\geq 1.$ By Lemma \ref{le1},
\bea\lbl{back_campus}
\lim_{p\to\infty}\zeta(p,d)=0
\eea
for each $d\geq 1$. The assertion \eqref{Upper_bound} implies that
\bea\lbl{639475}
P\Big(\bigcup_{i=1}^pA_pB_{i}\Big)
& \leq & P(A_p)\Big[\sum_{1\leq i_1 \leq p}P(B_{i_1})-\sum_{1\leq i_1< i_2\leq p}P(B_{i_1}B_{i_2})+\cdots-  \nonumber\\
&& \sum_{1\leq i_1<  \cdots < i_{2k} \leq p}P(B_{i_1}\cdots B_{i_{2k}})\Big]+ \Big[\sum_{d=1}^{2k}\zeta(p,d)\Big] + H(p, 2k+1)  \nonumber\\
&\leq & P(A_p)\cdot P\Big(\bigcup_{i=1}^pB_{i}\Big)+ \Big[\sum_{d=1}^{2k}\zeta(p,d)\Big] + H(p, 2k+1),
\eea
where the inclusion-exclusion formula is used again in the last inequality, that is,
\beaa
P\Big(\bigcup_{i=1}^pB_{i}\Big) &\geq & \sum_{1\leq i_1 \leq p}P(B_{i_1})-\sum_{1\leq i_1< i_2\leq p}P(B_{i_1}B_{i_2})+\cdots - \nonumber\\
&&~~~~~~~~~~~~~~~~ \sum_{1\leq i_1<  \cdots < i_{2k}\leq p}P(B_{i_1}\cdots B_{i_{2k}})
\eeaa
for all $k\geq 1$.
 By the definition of $l_p$ and \eqref{i2},
\beaa
 P\Big(\bigcup_{i=1}^pB_{i}\Big) \to 1-F(y)
\eeaa
as $p\to\infty$. By \eqref{i3}, $P(A_p)\to \Phi(x)$ as $p\to\infty.$ From \eqref{abci}, \eqref{back_campus} and \eqref{639475}, by fixing $k$ first and sending $p\to \infty$ we obtain that
\beaa
\limsup_{p\to\infty}P\Big(\frac{\z^\top\A\z-\tr(\A)}{\sigma_A}\leq x,\ \max_{1\le i\le p}|z_i|>l_p\Big)\leq \Phi(x)\cdot [1-F(y)] +\lim_{p\to\infty}H(p, 2k+1).
\eeaa
Now, by letting $k\to \infty$ and using \eqref{Maya1} we have
\bea\lbl{vskdnti}
\limsup_{p\to\infty}P\Big(\frac{\z^\top\A\z-\tr(\A)}{\sigma_A}\leq x,\ \max_{1\le i\le p}|z_i|>l_p\Big)\leq \Phi(x)\cdot [1-F(y)].
\eea
By applying the same argument to \eqref{Lower_bound}, we see that the counterpart of \eqref{639475} becomes
\beaa
P\Big(\bigcup_{i=1}^pA_pB_{i}\Big)
& \geq & P(A_p)\Big[\sum_{1\leq i_1 \leq p}P(B_{i_1})-\sum_{1\leq i_1< i_2\leq p}P(B_{i_1}B_{i_2})+\cdots + \nonumber\\
&& \sum_{1\leq i_1<  \cdots < i_{2k-1}\leq p}P(B_{i_1}\cdots B_{i_{2k-1}})\Big] + \Big[\sum_{d=1}^{2k-1}\zeta(p,d)\Big] - H(p, 2k)  \nonumber\\
&\geq & P(A_p)\cdot P\Big(\bigcup_{i=1}^pB_{i}\Big) + \Big[\sum_{d=1}^{2k-1}\zeta(p,d)\Big] - H(p, 2k).
\eeaa
where in the last step we use the inclusion-exclusion principle such that
\beaa
P\Big(\bigcup_{i=1}^pB_{i}\Big) &\leq & \sum_{1\leq i_1 \leq p}P(B_{i_1})-\sum_{1\leq i_1< i_2\leq p}P(B_{i_1}B_{i_2})+\cdots + \nonumber\\
&&~~~~~~~~~~~~~~~~ \sum_{1\leq i_1<  \cdots < i_{2k-1}\leq p}P(B_{i_1}\cdots B_{i_{2k-1}})
\eeaa
for all $k\geq 1$. Review \eqref{abci} and repeat the earlier procedure to see
\beaa
\liminf_{p\to\infty}P\Big(\frac{\z^\top\A\z-\tr(\A)}{\sigma_A}\leq x,\ \max_{1\le i\le p}|z_i|>l_p\Big)\geq \Phi(x)\cdot [1-F(y)]
\eeaa
by sending $p\to \infty$ and then sending $k\to\infty.$
This and \eqref{vskdnti} yield \eqref{th12}. The proof is completed. \hfill$\Box$

\subsection{Proof of Theorem \ref{th2}}

Taking the same procedure as Theorem 4 in \cite{tony2014two}, we have
\begin{align}\label{m1}
P\bigg(M_{{\O}^{1/2}}-2\log(p)+\log\log (p)\le x\bigg) \rightarrow \exp \bigg\{-\frac{1}{\sqrt{\pi}} \exp \bigg(-\frac{x}{2}\bigg)\bigg\}
\end{align}
where \begin{align*}
M_{{\O}^{1/2}}=\max_{1\le i \le p}  \nu_i^2,
\end{align*}
where $\nu_i=\frac{1}{\sqrt{n}}\sum_{k=1}^n \varepsilon_{ki}$. Let $  t=\sqrt{n}\bar{\X}$ we have
\begin{align*}
\left|||\hat{\O}^{1/2}  t||_{\infty}-||{\O}^{1/2}  t||_{\infty}\right|\le ||(\hat{\O}^{1/2}-{\O}^{1/2})  t||_{\infty}\le ||{\O}^{1/2}  t||_{\infty} ||(\hat{\O}^{1/2}{\O}^{-1/2}-\I_p)||_{L_1}
\end{align*}
By (\ref{m1}), we have $||{\O}^{1/2}  t||_{\infty}=O_p(\log (p))$. By condition (C3), we have $||(\hat{\O}^{1/2}{\O}^{-1/2}-\I_p)||_{L_1}=o_p(\log^{-1}(p))$. So $||\hat{\O}^{1/2}  t||_{\infty}-||{\O}^{1/2}  t||_{\infty}=o_p(1)$.
Here we obtain the result.

\subsection{Proof of Theorem \ref{th3}}
According to the proof of Theorem 3.1 in \cite{srivastava2009test}, we have
\begin{align*}
T_{SR}=\frac{n\bar{\X}^\top \D^{-1} \bar{\X}-p}{\sqrt{2\tr(\R^2)}}+o_p(1).
\end{align*}
And by the proof of Theorem \ref{th2}, we have $M_{\hat{\O}^{1/2}}=M_{\O^{1/2}}+o_p(1)$. So, by Lemma 7.10 in \cite{feng2022test}, we only need to show that
\begin{align}
P_{H_{0}}\bigg[\frac{n\bar{\X}^\top \D^{-1} \bar{\X}-p}{\sqrt{2\tr(\R^2)}}\le x,M_{{\O}^{1/2}}-2 \log (p)+\log \{\log (p)\} \leqslant y\bigg] \rightarrow \Phi(x)F(y).
\end{align}
By Theorem \ref{ind}, we only need to show that $\nu_i$ is independent sub-Gaussian random variables. Obviously,
\begin{align*}
E\bigg(\exp\bigg\{\frac{\lambda}{\sqrt{n}}\sum_{k=1}^n \varepsilon_{ki}\bigg\}\bigg)=E^n\bigg(e^{\frac{\lambda}{\sqrt{n}} \varepsilon_{ki}}\bigg)\le \bigg(E(1+\frac{\lambda}{\sqrt{n}} \varepsilon_{ki}+\frac{\lambda^2}{n} \varepsilon^2_{ki})+o(n^{-1})\bigg)^n\le e^{C\lambda^2}
\end{align*}
for large enough $n$ and some positive constant $C$. So we obtain the result.

\subsection{Proof of Theorem \ref{th4}}
Define $  \nu_{\mathcal{M}}$ be the sub-vector of $  \nu=(\nu_1,\cdots,\nu_p)$ corresponding to $i \in \mathcal{M}$. So does $  \nu_{\mathcal{M}^c}$. And let $\A_{\mathcal{M}},\A_{\mathcal{M}^c} $ be the sub-matrix of $\A$ corresponding to $\mathcal{M}$, $\mathcal{M}^c$, respectively. And $\A_{\mathcal{M}\mathcal{M}^c}$ is the sub-matrix between the vector $  \nu_{\mathcal{M}}$ and $  \nu_{\mathcal{M}^c}$.
According to the proof of Theorem 4.1 in \cite{srivastava2009test}, we have
\begin{align*}
T_{SR}=&\frac{n(\bar{\X} -   \mu)^\top \D^{-1} (\bar{\X} -   \mu)-p}{\sqrt{2\tr(\R^2)}}+\frac{  \delta^\top \D^{-1}  \delta}{p\sqrt{2\tr(\R^2)}}+o_p(1)\\
=&\frac{1}{\sqrt{2\tr(\R^2)}} (  \nu_{\mathcal{M}}^\top \A_{\mathcal{M}}  \nu_{\mathcal{M}})+\frac{2}{\sqrt{2\tr(\R^2)}} (  \nu_{\mathcal{M}}^\top \A_{\mathcal{M}\mathcal{M}^c}  \nu_{\mathcal{M}^c})+\frac{1}{\sqrt{2\tr(\R^2)}} (  \nu_{\mathcal{M}^c}^\top \A_{\mathcal{M}^{c}}  \nu_{\mathcal{M}^c} - p)\\
&+\frac{  \delta^\top \D^{-1}  \delta}{p\sqrt{2\tr(\R^2)}}+o_p(1)
\end{align*}
Additionally, by the proof of Lemma \ref{le1} and $m=o(p^{1/2})$, we have
\begin{align*}
P\bigg(  \nu_{\mathcal{M}}^\top \A_{\mathcal{M}\mathcal{M}^c}  \nu_{\mathcal{M}^c}\ge \epsilon\sqrt{2\tr(\R^2)} \bigg)\le K_{\epsilon}^m \exp(-c_{\epsilon}p^{1/2})\to 0\\
P\bigg(  \nu_{\mathcal{M}}^\top \A_{\mathcal{M}}  \nu_{\mathcal{M}}\ge \epsilon\sqrt{2\tr(\R^2)} \bigg)\le K_{\epsilon}^m \exp(-c_{\epsilon}p^{1/2})\to 0
\end{align*}
where $K_{\epsilon}$ and $C_{\epsilon}$ are two positive constant which dependent on $\epsilon$. So
\begin{align*}
T_{SR}
=&\frac{1}{\sqrt{2\tr(\R^2)}} (  \nu_{\mathcal{M}^c}^\top \A_{\mathcal{M}^c}  \nu_{\mathcal{M}^c}-p)+\frac{  \delta^\top \D^{-1}  \delta}{p\sqrt{2\tr(\R^2)}}+o_p(1)
\end{align*}
Similar to the proof of Theorem \ref{th2}, we have
$M_{\hat{\O}^{1/2}}=M_{\O^{1/2}}+o_p(1)$ where \begin{align*}
M_{{\O}^{1/2}}=\max_{1\le i \le p}  (\nu_i+\tilde{\mu}_i)^2=\max\{\max_{i \in \mathcal{M}}(\nu_i+\tilde{\mu}_i)^2, \max_{i \in \mathcal{M}^c}\nu_i^2\}.
\end{align*}
Under Condition (C1), we have $  \nu_{\mathcal{M}^c}^\top \A_{\mathcal{M}^c}  \nu_{\mathcal{M}^c}$ is independent of $\max_{i \in \mathcal{M}}(\nu_i+\tilde{\mu}_i)^2$. By \ref{th3},  $  \nu_{\mathcal{M}^c}^\top \A_{\mathcal{M}^c}  \nu_{\mathcal{M}^c}$ is asymptotically independent of $\max_{i \in \mathcal{M}^c}\nu_i^2$. So we obtain that $T_{SR}$ is asymptotically independent of $M_{\hat{\O}^{1/2}}$.

\subsection{Proof of Theorems \ref{th5}, \ref{th6} and \ref{th7}}

{\bf Proof of Theorem \ref{th5}} Similar to the proof of Theorem \ref{th2}, we have $W_{\hat{\O}^{1/2}}=W_{\O^{1/2}}+o_p(1)$. By Theorem 1 in \cite{tony2014two}, we have
\begin{align*}
P\bigg(W_{{\O}^{1/2}}-2\log(p)+\log\log (p)\le x\bigg)  \rightarrow \exp \bigg\{-\frac{1}{\sqrt{\pi}} \exp \bigg(-\frac{x}{2}\bigg)\bigg\}
\end{align*}
So we obtain the result.

\noindent{\bf Proof of Theorem \ref{th6}}
According to the proof of Theorem 1.1 in \cite{srivastava2013two}, we have
\begin{align}
T_{SKK}=\frac{1}{\sqrt{2\tr(\R^2)}}\bigg(  u^\top \A  u-p\bigg)+o_p(1)
\end{align}
where $\A$ is defined in condition (C5) and $  u=\sqrt{\frac{n_1n_2}{n_1+n_2}}\bigg(\frac{1}{n_1}\sum_{l=1}^{n_1}  \varepsilon_{1l} - \frac{1}{n_2}\sum_{l=1}^{n_2}  \varepsilon_{2l}\bigg)$. Similar to the proof of Theorem \ref{th3}, we can also prove that $  u$ is sub-Gaussian random variables. So by Theorem \ref{ind}, we can obtain the result.

\noindent{\bf Proof of Theorem \ref{th7}} The proof is similar to the proof of Theorem \ref{th4}. So we omit it here.

\th{Conflict of Interest} {\rm The authors declare no conflict of interest.}

\acknowledgements{\rm We thank the Editor, Associate Editor and the referees for their time and
comments. }

\end{document}